\documentclass[12pt]{article}

\newcommand{\blind}{0}

\usepackage{makecell}
\usepackage{multirow}
\usepackage{natbib}
\usepackage{graphicx,setspace}

\usepackage{amsmath,amsthm,amssymb}
\usepackage{xcolor}
\usepackage{rotating}
\usepackage{float} 
\usepackage{hyperref}
\hypersetup{
	colorlinks=true,
	linkcolor=blue,
	citecolor=blue,
	urlcolor=blue}
\usepackage{booktabs}
\usepackage{algorithm}
\usepackage{algpseudocode}
\bibpunct{(}{)}{;}{a}{,}{,}

\newtheorem{theorem}{Theorem}
\newtheorem{lemma}{Lemma}
\newtheorem{proposition}{Proposition}

\newtheorem{assumption}{Assumption}

\def\bb{\mathbf{b}}

\def\mA{{\mathcal A}}

\def\B{\mathbf{B}}

\def\ve{\varepsilon}

\def\beq{\begin{equation}}
	\def\eeq{\end{equation}}
\def\beqr{\begin{eqnarray}}
	\def\eeqr{\end{eqnarray}}
\def\beqrs{\begin{eqnarray*}}
	\def\eeqrs{\end{eqnarray*}}
\def\bet{\begin{theorem}}
	\def\eet{\end{theorem}}
\def\bel{\begin{lemma}}
	\def\eel{\end{lemma}}
\def\bep{\begin{proposition}}
	\def\eep{\end{proposition}}
\def\bg{\begin{figure}[tbph]\begin{center}}
		\def\eg{\end{center}\end{figure}}

\def\bc{\begin{center}}
	\def\ec{\end{center}}

\def\e{\mathbf{e}}

\newtheorem{remark}{Remark}

\def\wt{\widetilde}

\def\wh{\widehat}

\def\ol{\overline }

\def\E{\mathbb{E}}

\def\y{\mathbf{y}}
\def\A{\mathbf A}

\def\mC{\mathcal C}
\def\mN{\mathcal{N}}

\def\mG{\mathbb G}
\def\mR{\mathbb{R}}

\def\mL{\mathcal L}

\def\cH{\mathcal H}
\def\mH{\mathbb H}
\def\mM{\mathcal M}

\def\mF{\mathcal F}
\def\F{\mathbb F}
\def\cR{\mathcal{R}}

\def\mV{\mathcal{V}}
\def\mB{\mathcal{B}}
\def\cS{\mathcal{S}}

\def\mX{\mathbb{X}}
\def\cX{\mathcal{X}}

\def\bx{\mathbf{x}}

\def\u{\mathbf{u}}

\def\cov{\mbox{cov}}
\def\argmin{\mbox{argmin}}

\def\diag{\mbox{diag}}

\newcommand{\bgamma}{\boldsymbol{\gamma}}

\newcommand{\bxi}{\boldsymbol{\xi}}
\def\btheta{\boldsymbol{\theta}}
\def\bSigma{\boldsymbol{\Sigma}}
\def\bTheta{\boldsymbol{\Theta}}

\def\bPhi{\boldsymbol{\Phi}}
\def\bpsi{\boldsymbol{\psi}}

\def\bnu{\boldsymbol{\nu}}
\def\bmu{\boldsymbol{\mu}}

\def\bve{\boldsymbol{\varepsilon}}

\def\bfeta{\boldsymbol{\eta}}
\newcommand{\RNum}[1]{\uppercase\expandafter{\romannumeral #1\relax}}
\def\bX{\mathbf{X}}
\def\bx{\mathbf{x}}

\def\by{\mathbf{y}}

\def\bW{\mathbf{W}}

\def\bg{\mbox{\boldmath $g$}}
\def\bI{\mbox{\boldmath $I$}}

\def\zero{\mathbf{0}}
\def\defeq{\stackrel{\mathrm{def}}{=}}

\def\btheta{\boldsymbol{\theta}}
\def\bzeta{\boldsymbol{\zeta}}
\def\bSigma{\boldsymbol{\Sigma}}
\def\bTheta{\boldsymbol{\Theta}}

\def\bPhi{\boldsymbol{\Phi}}
\def\bpsi{\boldsymbol{\psi}}

\def\btau{\boldsymbol{\tau}}
\def\bnu{\boldsymbol{\nu}}
\def\bmu{\boldsymbol{\mu}}

\def\bfeta{\boldsymbol{\eta}}

\def\bvarphi{\boldsymbol{\varphi}}

\def\g{\mathbf{g}}

\def\bX{\mathbf{X}}
\def\bx{\mathbf{x}}

\def\by{\mathbf{y}}

\def\bW{\mathbf{W}}

\def\bg{\mbox{\boldmath $g$}}
\def\bI{\mbox{\boldmath $I$}}

\def\Z{\mathbf{Z}}

\def\W{\mathbf{W}}

\def\zero{\mathbf{0}}
\def\defeq{\stackrel{\mathrm{def}}{=}}

\def\x{\mathbf{x}}
\def\z{\mathbf{z}}
\def\v{\mathbf{v}}

\def\vr{\varrho}

\def\qic{\textup{QIC}}

\numberwithin{equation}{section}

\newcommand{\compressmath}{%
  \setlength{\abovedisplayskip}{1.5pt}
  \setlength{\belowdisplayskip}{1.5pt}
  \setlength{\abovedisplayshortskip}{1.5pt}
  \setlength{\belowdisplayshortskip}{1.5pt}
}

\def\boxit#1{\vbox{\hrule\hbox{\vrule\kern6pt\vbox{\kern6pt#1\kern6pt}\kern6pt\vrule}\hrule}}

\begin{document}

\def\spacingset#1{\renewcommand{\baselinestretch}%
{#1}\small\normalsize} \spacingset{1}

\if0\blind
{
\begin{center}
			{\bf\Large Two-way Homogeneity Pursuit for Quantile Network Vector Autoregression}\\
			\bigskip
			
			Wenyang Liu,$^{1,\dag}$ Ganggang Xu,$^{2,\dag}$ Jianqing Fan,$^{1, 3,*}$ and Xuening Zhu$^{4,*}$
			
			{\it $^1$ School of Data Science, Fudan University,
			$^2$ Department of Management Science, University of Miami, 
			$^3$ Department of Operations Research and Financial Engineering, Princeton University,
			and $^4$ School of Management, Fudan University.}

		\end{center}
			\begin{footnotetext}[1]			{\it $^{\dag}$Wenyang Liu and Ganggang Xu are joint first authors.
			$^*$Jianqing Fan and Xuening Zhu are joint corresponding authors. jqfan@princeton.edu (J. Fan); xueningzhu@fudan.edu.cn (X. Zhu).}
\end{footnotetext}

} \fi

\if1\blind
{
  \bigskip
  \bigskip
  \bigskip
  \begin{center}
    {\LARGE\bf Two-way Homogeneity Pursuit for Quantile Network Vector Autoregression}
\end{center}
  \medskip
} \fi

\vspace{-1.5em}
\begin{abstract}
We propose a two-way grouped network quantile (TGNQ) autoregression model for time series data observed on networks with substantial heterogeneity and complex directional interactions. Motivated by prior studies on network effects in directed networks, our model assigns each node two latent group memberships to flexibly capture asymmetric and heterogeneous interactions among users. These memberships, along with model parameters, can be consistently estimated using the proposed estimation procedure. As a result, the model performs node clustering and parameter estimation simultaneously, striking a balance between model flexibility and interpretability. We establish theoretical guarantees for the proposed method, showing that both group memberships and parameter estimators are consistent even when the number of groups is over-specified.
When the group numbers are correctly specified, the parameter estimators are asymptotically normal, enabling valid statistical inference.
In addition, we develop a quantile information criterion for the consistent selection of the number of groups. Simulation studies demonstrate strong finite-sample performance, and an application to Sina Weibo data illustrates the model's ability to uncover behavioral dynamics and user interaction patterns.

\end{abstract}

\noindent
{\it Keywords:}  Network data, Quantile vector autoregression, Two-way group structure, Two-way homogeneity pursuit.

\spacingset{1.9} 

\setlength{\abovedisplayskip}{5pt}
\setlength{\belowdisplayskip}{5pt}
\addtolength{\textheight}{.5in}
\vspace{-1.5em}
\section{Introduction}\label{section:intro}
 
\vspace{-0.5em}
Quantile regression~\citep{koenker1978regression} is a powerful statistical tool with wide-ranging applications, including economic growth analysis \citep{zhang2019quantile, chen2021quantile}, financial risk management \citep{hardle2016tenet, zhu2023quantile}, and air pollution studies \citep{chen2021joint}. In recent years, there has been growing interest in extending quantile regression to panel data settings \citep{kato2012asymptotics, galvao2016smoothed, he2023smoothed}. However, most existing approaches rely on assumptions of cross-sectional independence and homogeneity across individuals. For example, the quantile models of \cite{zhu2019network} and \cite{xu2024dynamic} impose common regression coefficients for all individuals. Such homogeneity assumptions may lead to model misspecification when individual-level heterogeneity is present. On the other hand, \cite{ando2023spatial} incorporate individual-specific spatial effects, which require estimating a large number of parameters in large-scale networks and may substantially reduce estimation efficiency. Consequently, effectively modeling individual heterogeneity within quantile vector autoregressive frameworks remains a challenging problem.

To strike a balance between model flexibility and estimation efficiency, a popular approach is to group the model coefficients.
For instance, \cite{ke2015homogeneity} develop a homogeneity pursuit approach for discovering clustered patterns among the model coefficients.
The group identification technique is also widely used for various panel data models; see \cite{bonhomme2015grouped}, \cite{bester2016grouped},
\cite{su2016identifying},
\cite{liu2020identification}, \cite{chen2023community} and \cite{chen2026group} for relevant literature.
Notably, for quantile regression models,
\cite{zhang2019quantile} develop a quantile regression-based clustering procedure for
identifying the subgroups of units in panel data.
\cite{gu2019panel} utilize a penalized estimation approach to discover the group structure imposed on the fixed effects of quantile panel models.
\cite{zhang2023nonparametric} propose a nonparametric quantile regression method for homogeneity pursuit with panel data.
However, they all assume independence among individuals and thus lack the ability to model  the potential
cross-sectional dependence.

To model cross-sectional dependence, we incorporate the observed network relationships among individuals into a quantile autoregression framework. Existing quantile autoregressive models often overlook the directional nature of interactions that commonly arise in network settings. \label{page:tw-moti-s}Recent studies on directed networks demonstrate that even the same user can exert markedly different effects on network dynamics depending on whether the user acts as a ``sender'' or a ``receiver'' of information as it propagates through the system \citep{choi2014co, rohe2016co, ji2016coauthorship, abbe2020entrywise, abbe2022lp, ke2023special}. For example, \cite{rohe2016co} distinguish between ``sending'' and ``receiving'' communities when designing a bi-clustering algorithm for directed networks. Similarly, \cite{aral2012identifying} characterize each user in a social network by two distinct traits: \textit{influence} (the ability to affect others) and \textit{susceptibility} (the extent to which the user is affected by others). The interaction between two users is jointly determined by the influence of the first and the susceptibility of the second, and their empirical analysis shows that these traits exhibit sharply different clustering patterns across users. \cite{zhou2024beyond} provide further empirical evidence that influence and susceptibility estimates yield superior predictive accuracy for identifying future superspreaders in a network.
  Motivated by these findings, we introduce a two-way group structure in which each user possesses two separate group memberships corresponding to these two traits. This stands in contrast to the one-way group structure proposed by \cite{zhu2025simultaneous}, where each network node is restricted to a single group identity. We emphasize that, relative to the one-way structure, adopting a two-way group specification brings substantial computational and theoretical challenges. A more detailed comparison of our modeling approach with existing group models and quantile regression methods is provided in Section A of the supplementary material.

In this work, we propose a two-way grouped network quantile (TGNQ) autoregressive model, where each network node is characterized by two latent group memberships to capture directional influence among different groups. By assuming that group memberships are shared across different quantiles, we can simultaneously identify memberships and estimate parameters at multiple quantile levels. We introduce an efficient algorithm for solving the highly non-convex loss function and establish asymptotic properties of the resulting estimators. Specifically, we derive clustering error rates and convergence rates of parameter estimators under the assumption of overspecified group numbers. Additionally, we develop a data-driven criterion for selecting the true group numbers with a probability approaching one. Given the correct identification of group numbers, we establish the asymptotic normality of estimated parameters at multiple quantile levels.

The rest of the paper is organized as follows.
Section \ref{section: model} presents the detailed model specifications and introduces the computational algorithms for the proposed approach.
Following this, Section \ref{section: properties} examines the theoretical properties of the proposed method, including estimation consistency and a method for consistently selecting group numbers.
In Section \ref{section: inference}, we establish the asymptotic normality of parameter estimators under the correct group number specification.
Subsequently, Section \ref{section: simu} presents Monte Carlo simulations to evaluate the finite-sample performance, while Section \ref{section: real} applies the method to real-world data.
The supplementary material provides additional details of the proposed algorithm, technical proofs, and supplementary experimental results.

\vspace{-1.1em}
\section{Homogeneity Pursuit with  Group Structures}\label{section: model}
\vspace{-0.5em}
\subsection{Model and Notation}\label{subsection: model}

\vspace{-0.5em}

Consider a network represented by an adjacency matrix $\A = (a_{ij})\in \{0,1\}^{N\times N}$, where $a_{ij}=1$ if an edge exists from the $i$th node to the $j$th node and $0$ otherwise.
Additionally, for all $i\in [N]$, $a_{ii} = 0$.\label{page:wii}
Let $\mN_i = \{j: a_{ij}\ne 0\}$ denote the neighboring set of node $i$, and let $n_i = |\mN_i|=\sum_{j=1}^N a_{ij}$ denote the out-degree of node $i$.
The row-normalized weighting matrix is then given by $\bW = (w_{ij})\in \mR^{N\times N}$, where $w_{ij} = a_{ij}/n_i$.
For each node, we observe a time series of continuous variables, $\{Y_{it}, 0\le t\le T\}$ for $1\le i\le N$.
Correspondingly, for each node, we collect a set of exogenous variables $\{\bx_{it}\in \mR^p:1\le t\le T\}$, where the dimension $p$ is fixed.
Denote the conditional quantile of $Y_{it}$ as $Q_{it}(\tau|\mF_{t-1})$, where $\mF_{t-1}$ is the $\sigma$-field generated by $\{(Y_{js}, \x_{j(s+1)}): s\le t-1, 1\le j\le N\}$.
Denote by $\btau_{ K}=\{\tau_1,\cdots,\tau_K\}$ a set of pre-specified quantiles, and we consider the following two-way grouped network quantile (TGNQ) autoregression model at the $\tau$th quantile
\beq
Q_{it}(\tau|\mF_{t-1}) = \sum_{j = 1}^N\theta_{g_i h_j}(\tau)w_{ij}Y_{j(t-1)} + \nu_{g_i}(\tau) Y_{i(t-1)} + \bx_{it}^\top\bgamma_{g_i}(\tau), \quad \tau\in\btau_K,
\label{gbnar}
\eeq
for any $1\le i\le N$, where
$g_i$ and $h_j$ are group memberships in the row and column, respectively.
In \eqref{gbnar}, we express the conditional quantile as a linear combination of three components. The first term represents a weighted average of the influence received from the followed nodes at the previous time point. In particular, $\theta_{g_i h_j}(\tau)$ quantifies the network effect from node $j$ to node $i$ at quantile level $\tau$, and this effect is assumed to be homogeneous within a two-way group membership $(g_i, h_j)$. Following \cite{aral2012identifying}, $g_i$ can be interpreted as the group (row-group) membership of user $i$ based on their susceptibility, while $h_j$ denotes the influence-based group (column-group) membership of node $j$. On the group level, $\theta_{gh}(\tau)$ reflects the network impact received by a member of susceptibility group $g$ from a member of influence group $h$, with $g \in [G]$ and $h \in [H]$. The term $\nu_{g_i}(\tau)$ then captures the autoregressive effect from the same node at the previous time point, while $\bgamma_{g_i}(\tau)$ denotes the corresponding covariate effect.
\label{page:fe}Extensions to include individual fixed effects in model~\eqref{gbnar} are discussed in Sections E.1 and E.2 of the supplementary material.

\vspace{-0.5em}
\begin{remark}
	We remark that the heterogeneous network effects have been noticed and modeled in recent literature.
	For instance, \cite{dou2016generalized} characterize the receiving
	network effects with different coefficients for each node $i$.
	This can be achieved by ignoring the column group in our model framework.
	\cite{zhu2019portal} model the influential power of each node $j$ with a regression coefficient associated with each node, and they identify important network nodes by a screening procedure.
	Similarly, within our framework, the influential power of node $j$ can also be characterized by setting $\theta_{g_ih_j}(\tau)$ to $\theta_{h_j}(\tau)$.
	A bi-directional network influence modeling framework with a network regression model is proposed by \cite{wu2022inward}, which encodes the network effect through a functional transformation of the nodes' covariates.
	Similar network autoregression models with one-way group structures are considered by \cite{zhu2025simultaneous} and \cite{fang2024group}, while they force the row and column group memberships to be identical.
\end{remark}

Let $\y_t=(Y_{1t},\cdots,Y_{Nt})^\top$. Then model~\eqref{gbnar} can be rewritten as follows
\compressmath
\begin{align*}
	Y_{it} =  {\bb_i}(\tau)^\top \y_{t-1} + \x_{it}^\top \bgamma_{g_i}(\tau) + \ve_{it}{(\tau)},  \quad \tau\in\btau_K,
\end{align*}
where $ {\bb_i}(\tau) = (w_{ij}\theta_{g_ih_j}(\tau): j\in [N])^\top+\e_i^{(N)} \nu_{g_i}(\tau) \in \mR^N$ with $\e_i^{(N)}$ being an $N$-dimensional unit vector with the $i$th element being one and the others being zero.
Here we define $\ve_{it}{(\tau)} = Y_{it} - {\bb_i}(\tau)^\top \y_{t-1} -  \x_{it}^\top \bgamma_{g_i}(\tau)$ for which it holds that  $P\left(\ve_{it}{(\tau)}<0|\mF_{t-1}\right) = \tau$. Consequently, the above model can be reorganized into the  matrix form
\begin{align}
	\label{model1}
	\y_t = \B(\tau) \y_{t-1} + \bmu_{x,t}(\tau) + \bve_{t}{(\tau)}, \quad \tau\in\btau_K,
\end{align}
where $\B(\tau) = ({\bb_i}(\tau): 1\le i\le N)^\top\in \mR^{N\times N}$,
$\bmu_{x,t}(\tau) = (\x_{it}^\top \bgamma_{g_i}(\tau): 1\le i\le N)^\top \in \mR^N$, and $\bve_t{(\tau)} = (\ve_{it}{(\tau)}: 1\le i\le N)^\top\in \mR^N$.

\begin{figure}[htpb!]
\centering
	\includegraphics[width=0.6\textwidth]{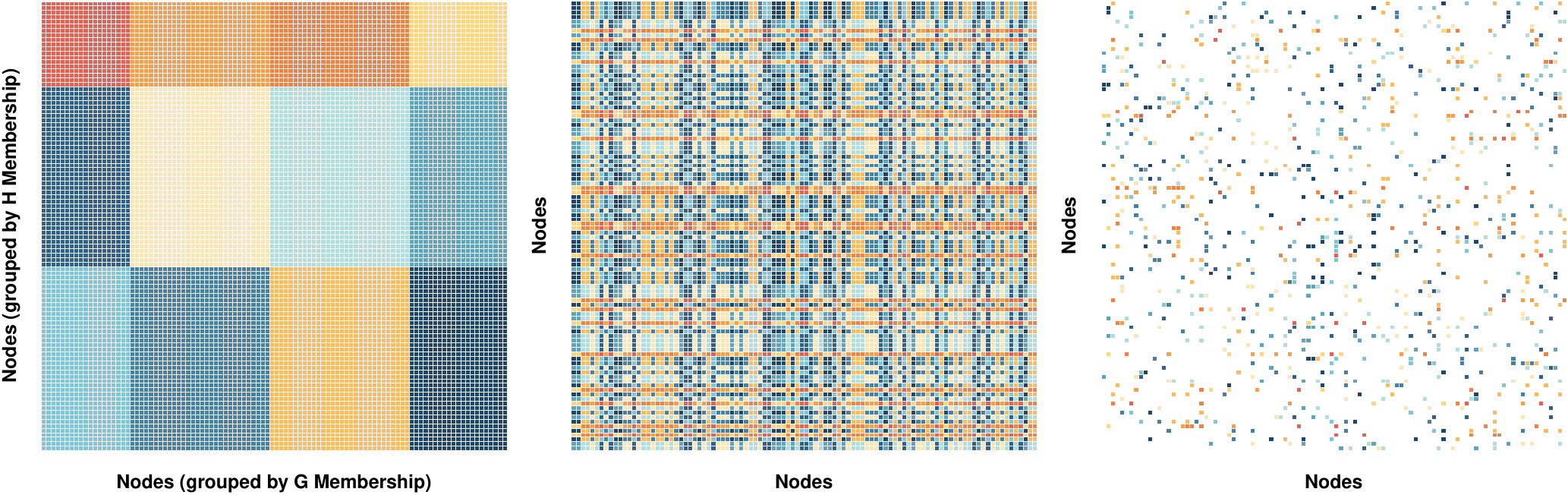}
	\caption{\small {A simulated example with $N=100$ nodes and $G=3$ and $H=4$ groups, respectively. Left panel: Row and column permuted $\bTheta(\tau)$; Middle panel: The original $\bTheta(\tau)$; right panel: Network sparsified  $\bTheta(\tau)$ (i.e., $\W\circ \bTheta(\tau)$).}}
	\label{fig:1}
\end{figure}

The proposed TGNQ model takes a vector autoregressive form \eqref{model1} with a structured transition matrix $\mathbf{B}(\tau)$, which can be expressed as follows
\compressmath
\[
\B(\tau)=\W\circ \bTheta(\tau) + \bnu(\tau), \quad \tau\in\btau_K,
\]
where $\circ$ denotes the Hadamard product,
$\boldsymbol{\Theta}(\tau) = (\theta_{g_ih_j}(\tau): i,j \in [N])\in \mR^{N\times N}$ represents the matrix of interactions,
and $\boldsymbol{\nu}(\tau) = \text{diag}\{\nu_{g_1}(\tau), \ldots, \nu_{g_N}(\tau)\}$. It is worth noting that by performing row and column permutations, it is possible to rearrange the rows and columns of ${\bTheta}(\tau)$ such that entries with the same group memberships are grouped together. Consequently, the permuted version of ${\bTheta}(\tau)$ exhibits a block matrix structure with ``rectangle-shaped'' blocks. See Figure~\ref{fig:1} for an illustration.

		From Figure~\ref{fig:1}, we can interpret the proposed model \eqref{model1} as follows: if the network is fully connected, the transition matrix $\mathbf{B}(\tau)$ can be permuted into a block matrix (except for the diagonal elements), which can be recovered by identifying the two-way group memberships. This process can be seen as a two-dimensional extension of the homogeneity pursuit approach \citep{ke2015homogeneity} to the VAR model. However, in practice, it is unrealistic for a network to be fully connected. Therefore, the interactions between nodes in the VAR model are further sparsified by imposing a network structure, leading to a sparse $\mathbf{B}(\tau)$.
		Let $\mG = (g_1,\cdots, g_N)^\top$ and $\mH = (h_1,\cdots, h_N)^\top$ denote the membership vectors for the row and column groups, respectively.
		To better understand the proposed model, we can further consider the following two special cases of the network effects $\theta_{g_ih_j}(\tau)$.
		
		\noindent
		{\bf The additive network effects.} We can further assume an additive form $\theta_{g_ih_j}(\tau) =\alpha_{g_i}(\tau)+\beta_{h_j}(\tau)$, and thus the transition matrix $\B(\tau)$ can be written as
		\compressmath
		\begin{align}
			\B(\tau) = \mA_{\mG}(\tau)\bW+\bW\mB_{\mH}(\tau)+ \mV_{\mG}(\tau), \quad \tau\in\btau_K,\label{eq:additive_network}
		\end{align}
		where $\mA_{\mG}(\tau)=\diag\{\alpha_{g_1}(\tau),\cdots,\alpha_{g_N}(\tau)\}$, $\mB_{\mH}(\tau)=\diag\{\beta_{h_1}(\tau),\cdots,\beta_{h_N}(\tau)\}$,
		and $\mV_{\mG}(\tau)=\diag\{\nu_{g_1}(\tau),\cdots,\nu_{g_N}(\tau)\}$.

		\noindent
		{\bf Multiplicative network effects.}  Alternatively, assume $\theta_{g_ih_j}(\tau) =\alpha_{g_i}(\tau)\times\beta_{h_j}(\tau)$, then
		\compressmath
		\begin{align}
			\B(\tau) = \mA_{\mG}(\tau)\bW\mB_{\mH}(\tau)+ \mV_{\mG}(\tau), \quad \tau\in\btau_K.\label{eq:multi_network}
		\end{align}
		
		In either case, the interaction between node $i$ and node $j$, i.e., $\theta_{g_ih_j}(\tau)$, can be viewed as the joint effect of the \textit{``susceptibility''} of node $i$,
		{i.e., $\alpha_{g_i}(\tau)$, and the  \textit{``influence''} of node $j$, i.e., $\beta_{h_j}(\tau)$.}
		Such additive or multiplicative decompositions of the network effects further improve the interpretability of the proposed TGNQ model and apply to many scenarios.

\vspace{-0.5em}
\begin{remark}\label{remark:iden}
To ensure parameter identifiability, we need to impose identification restrictions for the additive and multiplicative models, respectively.
For the additive model, we set $\alpha_1(\tau) = 0$ for all $\tau \in \btau_K$,
thereby fixing the baseline row group effect and ensuring unique parameter estimates.
Similarly, for the multiplicative model, we set $\alpha_1(\tau) = 1$ for all $\tau \in \btau_K$ to eliminate multiplicative ambiguity.
Note that the general model formulation does not require such constraints, as its structure inherently ensures identifiability.
\end{remark}

\vspace{-1.1em}

		\subsection{Model Estimation}\label{sec:model_est}
		\vspace{-0.5em}
		Let $\btheta = (\theta_{gh}(\tau_k): g\in [G], h\in [H],1\le k \le K)^{\top}\in \mR^{KGH}$,
		$\bnu = (\nu_g(\tau_k): g\in [G],1\le k \le K)^\top\in \mR^{KG}$, and
		$ \bgamma =  (\bgamma_g(\tau_k)^\top: g\in [G], 1\le k \le K)^\top\in \mR^{ KGp}$ collect all model parameters for  pre-specified
		group numbers $G$ and $H$.
		For convenience, we denote $\bpsi = \{\btheta, \bnu, \bgamma\}$.
		Recall that $\mG = (g_i: i\in [N])^\top$ and $\mH = (h_j: j\in [N])^\top$ are the row and the column group membership vectors,
		respectively.
		Subsequently, the unknown parameters and memberships can be estimated by $(\wh \bpsi, \wh \mG, \wh \mH)$, which minimizes the following multiple quantile loss function
		\compressmath
		\begin{align}
			&\mL\left(\bpsi, \mG, \mH\right)\defeq \frac{1}{K}\sum_{k = 1}^K \mL_{\tau_k}(\bpsi, \mG, \mH)\nonumber\\
			& {\defeq \frac{1}{NTK}\sum_{k=1}^K\sum_{i=1}^N \sum_{t=1}^T}\rho_{\tau_k} \Big(Y_{it} - \sum_{j = 1}^N\theta_{g_i h_j}(\tau_k)w_{ij}Y_{j(t-1)} - \nu_{g_i}(\tau_k) Y_{i(t-1)}- \bx_{it}^\top\bgamma_{g_i}(\tau_k)\Big),\label{eq:loss_func}
		\end{align}
		where $
		\rho_\tau(u) = u\{\tau - I(u<0)\}$ is the check function for the $\tau$th quantile.
		With unknown $\mG$ and $\mH$, the loss function~\eqref{eq:loss_func} is highly nonconvex, rendering its minimization a challenging mixed integer optimization problem, for which we have devised the following algorithms.

\vspace{-1.1em}
		\subsubsection{The Vanilla Algorithm}
		\vspace{-0.5em}
		
		We begin by introducing a simple algorithm that iteratively updates the model parameters and group memberships through coordinate descent. Let $\cR_g = \{i:g_i = g\}$ and $\mC_h = \{j: h_j = h\}$ for $g\in[G]$, $h\in[H]$.
		Define $\y_t^g =  (Y_{it}: i \in \cR_g)^\top\in \mR^{|\cR_g|}$,
		$\ol \y_t^{(g,h)} = (\sum_{j\in \mC_h} w_{ij}Y_{jt}: i\in \cR_g)^\top\in \mR^{|\cR_g|}$,
		and $Q_{\y_t^g}(\tau_k| \mF_{t-1}) = \{Q_{it}(\tau_k|\mF_{t-1}): i \in \cR_g\}^\top \in \mR^{|\cR_g|}$ as the $\tau_k$th quantile vector for $\y_t^g$.
		Then, model~\eqref{gbnar} can be written as
		\compressmath
		\begin{equation}
			\label{eq:d}
			Q_{\y_{t}^g}(\tau_k|\mF_{t-1})=\sum_h\ol\y_{t-1}^{(g,h)}
			\theta_{gh}(\tau_k)+ \nu_g(\tau_k)\y_{t-1}^g+ \bX_t^{g} \bgamma_g(\tau_k)\defeq \Z_{t-1}^g\bxi_g(\tau_k),
		\end{equation}
		where
		$\bX_t^g = (\x_{it}: i\in \cR_g)^\top\in \mR^{|\cR_g|\times p}$,
		$\Z_{t-1}^g = (\ol\Z_{t-1}^g,\y_{t-1}^g,  \bX_t^{g})\in \mR^{|\cR_g|\times (H+p+1)}$
		with $\ol\Z_{t-1}^g = (\ol\y_{t-1}^{(g,h)}: h\in [H])\in \mR^{|\cR_g|\times H}$, and $\bxi_g(\tau_k) = (\theta_{g1}(\tau_k), \cdots \theta_{gH}(\tau_k), \nu_g(\tau_k), \bgamma_g^\top(\tau_k))^\top\in \mR^{H+p+1}$.

\vspace{-1em}

		\begingroup
		\singlespacing
		\begin{algorithm}[ht!]
				\footnotesize
			\caption{ The Vanilla Algorithm}\label{algo}
			Denote $\wh\bpsi^{(l)}$, $\wh\mG^{(l)}$ and  $\wh\mH^{(l)}$ as the estimators obtained after the $l$th iteration.
			\vspace{-5pt}
			\begin{enumerate}
				\item[]\text{\hskip -2em} {\bf Step I}: fixing $\mG=\wh\mG^{(l)}$ and $\mH=\wh\mH^{(l)}$, update $\wh\bpsi^{(l)}$ by $\wh\bpsi^{(l+1)}$, whose components are estimated as
				$\wh \bxi_g^{(l+1)}(\tau_k) = \arg\min_{\bxi_g}{\sum_{t=1}^T}\sum_{i\in \cR_g}\rho_{\tau_k}(Y_{it} - \z_{i,t-1}^{g\top}\bxi_g)$ for $g \in [G],\tau_k\in \btau_K$, where $\z_{i,t-1}^{g{\top}}$ is the $i$th row of $\Z_{t-1}^g$.
	 			\vspace{-5pt}
				\item[]\text{\hskip -2em} {\bf Step II}: fixing $\bpsi=\wh\bpsi^{(l+1)}$ and $\mH=\wh\mH^{(l)}$, update $\wh\mG^{(l)}$ by $\wh\mG^{(l+1)}=(\wh g_1^{(l+1)},\cdots,\wh g_N^{(l+1)})^\top$,  where $\wh g_i^{(l+1)}$ minimizes the following loss function (with respect to $g_i$):
				\compressmath
				\beq
\sum_{k  =1}^K\sum_{t=1}^T\rho_{\tau_k}\{Y_{it} - \sum_{j = 1}^N\wh \theta_{g_i \wh h_j^{(l)}}^{(l+1)}(\tau_k)w_{ij}Y_{j(t-1)} - \wh\nu_{g_i}^{(l+1)}(\tau_k) Y_{i(t-1)}- \bx_{it}^\top\wh\bgamma_{g_i}^{(l+1)}(\tau_k)\}.\label{eq:update_g}
\eeq
				\vspace{-5pt}
				\item[]\text{\hskip -2em} {\bf Step III}: fixing $\bpsi=\wh\bpsi^{(l+1)}$ and $\mG=\wh\mG^{(l+1)}$, use the coordinate-descent algorithm to update $\wh\mH^{(l)}$ by $\wh\mH^{(l+1)}=(\wh h_1^{(l+1)},\cdots,\wh h_N^{(l+1)})^\top$,  where $\wh h_j^{(l+1)}$ minimizes the following loss function (with respect to $h_j$):
				\begin{align}
					\sum_{k=1}^K\sum_{i=1}^N \sum_{t=1}^T
					\rho_{\tau_k}\Big(&Y_{it}
					- \sum_{j': j' \neq j}^N\wh\theta_{\wh g_i^{(l+1)} \wt h_{j'}}^{(l+1)}(\tau_k)w_{ij'}Y_{j'(t-1)}
					\nonumber\\
					&- \wh\theta_{\wh g_i^{(l+1)} h_j}^{(l+1)}(\tau_k)w_{ij}Y_{j(t-1)}
					- \wh\nu_{\wh g_i^{(l+1)}}^{(l+1)}(\tau_k) Y_{i(t-1)}- \bx_{it}^\top\wh\bgamma_{\wh g_i^{(l+1)}}^{(l+1)}(\tau_k)\Big),\label{eq:update_h}
				\end{align}
				where $\wt h_{j'}=\wh h_{j'}^{(l+1)}$ if $j'<j$ and $\wt h_{j'}=\wh h_{j'}^{(l)}$ if $j'>j$. Repeat Step III for several rounds until no $\wh h_j^{(l+1)}$ can be changed.
				\vspace{-5pt}
				\item[]\text{\hskip -2em} {\bf Repeat} Steps I-III until the algorithm converges.
			\end{enumerate}
		\end{algorithm}
		\endgroup
		
		\vspace{-1em}
		
	Algorithm~\ref{algo} outlines the vanilla algorithm for estimating the general TGNQ model~\eqref{gbnar}.
	To implement this algorithm effectively, a suitable initialization procedure is required, which is detailed in Section B.1 of the supplementary material.
	However, the estimation procedures for the additive model~\eqref{eq:additive_network} and the multiplicative model~\eqref{eq:multi_network} differ slightly from Algorithm~\ref{algo}, as we can further exploit their distinct structures. Detailed explanations are provided in Section B.3 of the supplementary material.
	In addition, an alternative initialization based on graph Laplacian regularization for the additive model is developed and numerically evaluated in Section B.2 of the supplementary material.

\vspace{-1.1em}
	\subsubsection{The Enhanced Algorithm}\label{sec:enhanced}

		\vspace{-0.5em}
		The Vanilla Algorithm~\ref{algo} mirrors the algorithm employed in~\cite{zhu2025simultaneous}, which proves particularly effective when each node assumes only one group membership. However, our empirical observations reveal that this method encounters difficulty in accurately determining the column membership $\mH$ because the coordinate descent algorithm often gets trapped in local minima. To tackle this challenge, we propose an enhanced algorithm specifically tailored to facilitate escaping from local minima while estimating $\mH$.
		
		After obtaining $\wh\bpsi$, $\wh\mG$, and $\wh\mH$ from Algorithm~\ref{algo}, we continue to find a good set of proposals for $\mH$ to escape from the possible local minima $\wh\mH$ by dividing the nodes into two sets: the active set $\mA$ and the inactive set $\mA^c$.  If $j\in\mA^c$, then $\wh h_j$ is considered reliable and need not be updated.
		In the beginning, we let the active set $\mA_{(0)} = \{1,\cdots, N\}$.
		
		Next,
		in the $r$th round ($r = 1,2,\cdots$), we calculate the following
		\begin{equation}
			\label{eq:h2}
			\footnotesize
			(\wt h_{ij}: j\in \mA_{(r-1)}\cap\mN_i)=\underset{(h_j: j\in \mA_{(r-1)}\cap\mN_i)}{\argmin} \sum_{k=1}^K \sum_{t=1}^T\rho_{\tau_k} \Big(r_{it}(\tau_k)- \sum_{j \in \mA_{(r-1)}\cap \mN_i}\wh\theta_{\wh g_i h_j}(\tau_k)w_{ij}Y_{j(t-1)} \Big),
		\end{equation}
		for each $1\le i\le N$, where $r_{it}(\tau_k)=Y_{it} -  \sum_{j \in \mA_{(r-1)}^c\cap \mN_i}\wh\theta_{\wh g_i \wh h_j}(\tau_k)w_{ij}Y_{j(t-1)} - \wh\nu_{\wh g_i}(\tau_k) Y_{i(t-1)}- \bx_{it}^\top\wh\bgamma_{\wh g_i}(\tau_k)$
		and the estimators $\{\wh \bpsi, \wh \mG\}$ are fixed. Given that the cardinality of $\mA_{(r-1)}\cap\mN_i$ is typically not large, solving the above minimization problem through enumeration is straightforward. However, in cases where $|\mA_{(r-1)}\cap\mN_i|$ is relatively large, we resort to using the discrete optimization routine provided by the R package {\sf CEoptim}~\citep{benham2017ceoptim} to solve it.
		Then, we update an estimator $\wt\mH_{(r)}=(\wt h_1,\cdots,\wt h_N)^\top$ for estimating $\mH$ based on the $\{\wt h_{ij}\}$ obtained in \eqref{eq:h2}.
		Specifically, for $j\in\mA_{(r-1)}^c$, $\wt h_j$ is set to $\wh h_j$,
		while for $j\in\mA_{(r-1)}$, $\wt h_j$ is randomly drawn from the vector $(\wt h_{ij}: i \in \F_j)$, where $\F_j = \{i: a_{ij}=1\}$ represents the set of followers for node $j$.
		When $\wh\bpsi$ and $\wh\mG$ are close to the truth, one naturally expects that $\wt\mH$ should also be reasonably close to the truth. Therefore, if $\wt h_j$ coincides with $\wh h_j$ for $j\in \mA_{(r-1)}$, there would be more confidence in their accuracy, prompting the removal of node $j$ from the active set. Consequently, we update the active set as $\mA_{(r)}=\{j\in[N]: \wh h_j \ne \wt h_j\}$.

		Subsequently, we randomly partition the nodes in $\mA_{(r)}$ into $\mA_{(r,1)}$ and $\mA_{(r,2)}$, and denote $\wt \mH_{(r,s)}$ ($s = 1,2$) as the column membership vector by
		replacing $\wh h_j$ with $\wt h_j$ for $j \in \mA_{(r,s)}$ in $\wh\mH$.
		Next, we check if $\wt \mH_{(r,s)}$ ($s = 1,2$) can improve the objective function by using $\wh\bpsi$, $\wh\mG$, and $\wt \mH_{(r,s)}$ as the initial inputs for Vanilla Algorithm~\ref{algo}.
		If the objective function fails to improve,
		we can resample $\wt\mH_{(r)}$ or update the active set $\mA_{(r-1)}$ for more attempts until the algorithm converges. The specific iterative steps are summarized in Algorithm \ref{algo2}.
			
			\vspace{-1.2em}
			
			\begingroup
		\singlespacing
		\begin{algorithm}[ht!]
		\footnotesize
			\caption{The Enhanced Algorithm}\label{algo2}
			\begin{enumerate}
				\item[]\text{\hskip -2.5em} {\bf Step I} {\sc(Initialization)}  Set $r= 0 $  and the active set $\mA_{(0)}=\{1,\cdots,N\}$.
				Run Algorithm~\ref{algo} and denote the corresponding output as $\wh\bpsi_{(r)}$, $\wh\mG_{(r)}$ and  $\wh\mH_{(r)}$.
					
				 \vspace{-7pt}
				\item[]\text{\hskip -2.5em} {\bf Step II} {\sc(Draw proposals from updated Active Set)} Obtain $\wt h_{ij}, i\in [N], j\in \mA_{(r)} \cap \mN_i$ using~\eqref{eq:h2}. Set $r=r+1$.
				\vspace{-7pt}
				\begin{itemize}
					\item {\bf If} $\wt h_{ij} = \wh h_{(r)j}$ for all $i\neq j \in [N]$ {\bf or} $r>r_{\max}$, {\bf end} the algorithm and return
					$\{\wh\bpsi_{(r-1)},\wh\mG_{(r-1)},\wh\mH_{(r-1)}\}$.
				\end{itemize}
				
				\vspace{-7pt}
				\item[]\text{\hskip -2.5em} {\bf Step III} {\sc(Improving the Objective Function)}\\
				(3.1) Obtain $\wt\mH_{(r)}$ by randomly drawing $\wt h_j$ for $j\in\mA_{(r-1)}$ from the vector $(\wt h_{ij}: i \in \F_j)$ and update $\mA_{(r)} = \{j: \wt h_j \neq \wh h_{(r)j}\}$.
				\\
				(3.2) Uniformly split $\mA_{(r)}$ to obtain $\wt \mH_{(r,1)}$ and $\wt \mH_{(r,2)}$. \\
				(3.3) Use $\wh\bpsi_{(r-1)}$, $\wh\mG_{(r-1)}$, and $\wt \mH_{(r,s)}, s =1,2$ as the initial inputs for Algorithm~\ref{algo}.\\
				(3.4) Repeat Steps (3.1)--(3.3). {\bf If} the objective fails to improve for $10$ consecutive times,  go to {\bf Step II}; {\bf else} go to {\bf Step I}.

				\vspace{-5pt}
				\item[]\text{\hskip -2.5em} {\bf Output} $\{\wh\bpsi_{(r)},\wh\mG_{(r)},\wh\mH_{(r)}\}$.
			\end{enumerate}
		\end{algorithm}
		\endgroup	
		As $r$ increases in Step II of Algorithm~\ref{algo2}, the active set $\mA_{(r)}$ diminishes,
indicating fewer memberships in $\wh\mH$ need to be altered, which potentially accelerates convergence.
		In our numerical experiments, the enhanced Algorithm~\ref{algo2} shows excellent results, consistently achieving accurate identification of true group memberships with high probability as $T$ increases.
		However, providing theoretical justification for Algorithm~\ref{algo2} poses a significant challenge and remains an intriguing avenue for future research.

Lastly, we note that the computational cost of Algorithm \ref{algo2}
is relatively heavier than that of Algorithm \ref{algo}.\label{page:algo_time}
The main reason that Algorithm \ref{algo2} repeatedly invokes Algorithm \ref{algo} to escape from local minima when estimating the column memberships $\mH$.
Our simulations indicate that the computation time increases much more rapidly with $N$ than with $T$,
since a larger $N$ tends to require more invocations for convergence.
In Section B.6 of the supplementary material, we examine how the computation time scales as $N$ or $T$ increases, and discuss strategies for improving computational efficiency.
In particular, we consider the use of modern techniques such as the convolution-type smoothed quantile regression ({\sf conquer}) method proposed by \citet{he2023smoothed} to improve computational scalability as network datasets continue to grow in size.
We further report wall-clock time comparisons between the standard quantile regression solver and the {\sf conquer} method, as well as its parallelized variant, which together substantially alleviate the computational burden.

\vspace{-1.1em}
		\section{Theoretical Properties}\label{section: properties}
	\vspace{-0.5em}	
		
		In this section, we investigate the theoretical properties of the TGNQ parameter estimators.
		Denote the true parameter as $\boldsymbol{\psi}^0 = \{\boldsymbol{\theta}^0, \boldsymbol{\nu}^0, \boldsymbol{\gamma}^0\}$ and the true numbers of row and column groups as $G_0$ and $H_0$. We first present a set of technical conditions in the following section.
		
		\vspace{-1em}
		\subsection{Technical Conditions}\label{section:condition}
		\vspace{-0.5em}
		
		To establish the theoretical properties, we require the following technical conditions.
		
		\vspace{-1em}
		\begin{assumption}\label{assum:f_bound}
			{\sc (Conditional Density)}
			Assume that for each $i$, the stochastic process $\bzeta_{it} = (\by_{t-1, \mN_i}^\top,Y_{i,t-1}, \x_{it}^\top)^\top$, $t=1,\cdots, T$, is strictly stationary over $t$, where $\by_{t-1, \mN_i} = (Y_{j,t-1}: j \in \mN_i)^\top \in \mR^{n_i}$.
			Denote
			the conditional density of $\ve_{it}(\tau)$ given $\bzeta_{it}$ as $f_{i,\tau}(u|\bzeta_{it})$,
			and we further assume that there exists a constant $\ol f$ such that $\max_{ i,k}\sup_{u,\bzeta}|f_{i,\tau_k}(u|\bzeta)|<\ol f$ and $\max_{ i,k}\sup_{u,\bzeta}|f_{i,\tau_k}'(u|\bzeta)|<\ol f$, where $f_{i,\tau_k}'(u|\bzeta)$ is the first order derivative
			of $f_{i,\tau_k}(u|\bzeta)$.
		\end{assumption}
		
		\vspace{-1em}
		\begin{assumption}\label{assum:Theta_finite}
			
			{\sc (Parameter Space)}
			Assume that $\max\{\|\btheta\|_{\max}, \|\bnu\|_{\max}, \|\bgamma\|_{\max}\} <c$, where $c$ is a finite constant.
		\end{assumption}
		
		\vspace{-1em}
		\begin{assumption}\label{assum:tau_min}
			{\sc (Moment Conditions)}
			Define the matrix
			$\bSigma_{i}(\tau) = \E\left\{f_{i,\tau}(0|\bzeta_{it})\bzeta_{it}\bzeta_{it}^\top\right\}$.
			Assume that (a) $\max_{i,t}\sup_{\|\u\| = 1} \E(|\bzeta_{it}^\top\u|^3) = O(1)$, (b) $\lambda_{\max}(\E(\bzeta_{it}\bzeta_{it}^\top))\le\sigma_{\max}$, and that (c) $ \min_{i,k}\lambda_{\min}(\bSigma_i(\tau_k))\ge \sigma_{\min}$, where $\sigma_{\min}$ and $\sigma_{\max}$ are finite positive constants.
		\end{assumption}
		
\vspace{-1em}
		\begin{assumption}\label{assum:group_sep}
			{\sc (Group Differences)}
			Let $c_{G,NT}$ and $c_{H,NT}$ be sequences with $c_{G,NT},c_{H,NT}\to 0$ as $(N,T)\to\infty$.
			Assume that
			(a) $\min\limits_{g\ne g'\in[G_0]}\{K^{-1}\sum_{k=1}^{K}(|\nu_{g}^0(\tau_k) - \nu_{g'}^0(\tau_k)|^2$ +
			$\|\bgamma_{g}^0(\tau_k) - \bgamma_{g'}^0(\tau_k)\|^2)\}\geq c_{G,NT}$;
and (b) $\min_{h_0\ne h_0'\in[H_0], g_0\in [G_0]}	(K^{-1}\sum_{k=1}^{K}|\theta_{g_0h_0}^0(\tau_k) - \theta_{g_0h_0'}^0(\tau_k)|^2)\geq  c_{H,NT}
			$
		\end{assumption}
		
\vspace{-1em}
\begin{assumption}
{\sc (Network Structure)}\label{assum:net}
Define $q_j = \sum_i w_{ij}^2$.
Let $c_{w,NT}$ be a sequence such that $c_{w,NT}\to 0$ as $(N,T)\to \infty$.
Assume that there exists a set $\mM_H\subset [N]$ such that $|\mM_H^c|/N\to 0$ and $\min_{j\in \mM_H} q_j \geq c_{w,NT}$.
\end{assumption}

		\vspace{-1em}
		\begin{assumption}\label{assum:c_pi}
			{\sc (Group Ratios)}
			Let $\pi_{g_0}^{(1)} = \sum_{i}I(g_i^0 = g_0)/N$,
			$\pi_{h_0}^{(2)} = \sum_{j}I(h_j^0 = h_0)/N$, and
			$\pi_{g_0h_0} = N^{-1}\sum_{i,j}I(g_i^0 = g_0, h_j^0 = h_0)w_{ij}^2$.
			Assume that $\min_{g_0\in [G_0], h_0\in [H_0]} \{\pi_{g_0}^{(1)},
			\pi_{h_0}^{(2)}, \pi_{g_0h_0}\}\ge c_\pi$, where $c_\pi$ is a positive constant.
		\end{assumption}
		
		\vspace{-1em}
		\begin{assumption}\label{assum:beta_mixing}
			{\sc (Distribution Conditions)}
			Let $\z_{t}\defeq (\y_t^\top,\x_{1t}^\top, \x_{2t}^\top,\cdots, \x_{Nt}^\top)^\top$ and assume the following conditions hold. (a) The process $\{\z_t\}$ is geometrically $\beta$-mixing; that is,
			there exist constants $c>0$ and $\gamma_1>0$ such that
			$
			\beta(n)\le 2\exp(-cn^{\gamma_1})
			$
			for any $n>1$, where $\beta(n)$ denotes  the $\beta$-mixing coefficient
			for the $\sigma$-fields associated with $\{\z_s: s\le t\}$ and $\{\z_{s}: s\ge t+n\}$. (b) The random vector $\z_t$ follows a sub-Weibull$(\gamma_2)$ distribution in the sense that there exists a constant $K_z$ such that $\|\z_t\|_{\phi_{\gamma_2}}\le K_z<\infty$ for $t\in [T]$, where
			the sub-Weibull$(\gamma)$ norm for a random variable $X$
			is defined as $\|X\|_{\phi_{\gamma}} = \sup_{p\ge 1}(\E|X|^p)^{1/p}p^{-1/\gamma}$
			and
			$\|\z\|_{\phi_{\gamma}} = \sup_{\v\in \mR^n, \|\v\| = 1} \|\v^\top \z\|_{\phi_{\gamma}}$.
		\end{assumption}
		
		\vspace{-1em}
		\begin{assumption}\label{assum:sample_size}
		{\sc (Sample Sizes)}
			 (a) Let $\gamma = (\gamma_1^{-1}+\gamma_2^{-1})^{-1}<1$ and  $T^{\gamma_3} \gg n_{\max}\log(NK)$ with $\gamma_3 = (\gamma_1^{-1}+3\gamma_2^{-1})^{-1}$, where $n_{\max} \defeq \max_{i\in [N]}n_i$ is the maximum out-degree of nodes.
			 (b) Assume that, as $(N,T)\to \infty$
			 \begin{align*}
\min\{c_{G, NT}, c_{H, NT}\} \gg \sqrt{\frac{\ol n\log(NK)}{T}}+\frac{(\log(NKT))^{1/\gamma}}{T}, \quad c_{G, NT}c_{H, NT}{c_{w,NT}} \gg \frac{\ol n \log(NK)}{T}.
\end{align*}
\end{assumption}
\label{page:R2Q2}All assumptions presented above are widely adopted in the relevant literature. 
Assumptions~\ref{assum:f_bound} and \ref{assum:beta_mixing} are high-level conditions 
imposed to ensure model stationarity, weak temporal dependence, and well-behaved tails 
without specifying a particular data-generating mechanism, thereby maximizing the applicability of the proposed methodology. 
In Section~E.3.4 of the supplementary material, we provide low-level sufficient conditions under which the quantile autoregressive model \citep{koenker2006quantile,zhu2023quantile} in \eqref{eq:simu_model} satisfies Assumptions~\ref{assum:f_bound} and \ref{assum:beta_mixing}. Sensitivity analyses related to these assumptions are presented in Section~G.1 of the supplementary material.
\label{page:coro1} Assumption \ref{assum:group_sep} is an identifiability condition that essentially requires gaps with lower bounds $c_{G,NT}$ and $c_{H,NT}$ between different groups, ensuring distinguishability among groups.
Here, $c_{G,NT}$ and $c_{H,NT}$ are positive sequences that may depend on $N$ and $T$, allowing the gaps to diminish as sample sizes increase, with their rates controlled by Assumption \ref{assum:sample_size}(b).
Assumption \ref{assum:group_sep}(a), applied to the row groups, extends a similar condition in \cite{liu2020identification} by no longer requiring a constant gap, and Assumption \ref{assum:group_sep}(b) imposes an analogous condition on the column groups.
Furthermore, Assumption \ref{assum:net} characterizes the network structure through the in-degree information $q_j = \sum_i w_{ij}^2$, which we refer to as the {\sf signal strength} of node $j$. 
Intuitively, $q_j$ needs to be sufficiently large to ensure the correct identification of node $j$'s column group membership. 
Accordingly, we refer to $\mathcal{M}_H$ as the {\sf informative set}, which collects nodes with in-degree information sufficient for consistent clustering. 
The condition $|\mathcal{M}_H^c|/N \to 0$ then requires that the proportion of marginal nodes, whose column memberships may not be correctly identified, becomes asymptotically negligible.
More detailed discussions of these assumptions, including sufficient conditions and verification of related properties, are provided in Section E.3 of the supplementary material, where intuitive examples for Assumptions \ref{assum:net} are also given.

		\vspace{-1.1em}
		\subsection{Estimation Consistency When $G\ge G_0$ and $H\ge H_0$}\label{section: consistency}
		\vspace{-0.5em}
		
		We first establish the estimation consistency when the group numbers are possibly overspecified, i.e., $G\ge G_0$ and $H\ge H_0$. The following theorem provides the overall convergence rate of all model parameters, including group memberships.
		
		\bet\label{thm:consistency}
		Under Assumptions \ref{assum:f_bound}--\ref{assum:sample_size}, if $G\ge G_0$ and $H\ge H_0$, then we have that
		\compressmath
		\begin{align}
			\sup_{k\in[K]}\frac{1}{N}\sum_{i= 1}^N \Big\{\sum_{j\in \mN_i}w_{ij}^2|\wh \theta_{\wh g_i\wh h_j}(\tau_k)
			- \theta_{g_i^0h_j^0}^0(\tau_k)|^2 + |\wh \nu_{\wh g_i}(\tau_k) &- \nu_{g_i^0}^0(\tau_k)|^2
			\nonumber\\
			+ \big\|\wh \bgamma_{\wh g_i}(\tau_k) - \bgamma_{g_i^0}^0(\tau_k)\big\|^2\Big\}
			= O_p\left(\frac{\ol n \log(NK)}{T}\right),\label{eq:d_w_conv}
		\end{align}
		where $\ol n = N^{-1}\sum_i n_i$ represents the average out-degree of nodes.
		\eet

		\vspace{-0.5em}
		The proof of Theorem \ref{thm:consistency} is given in Section C.2 of the supplementary material.
\label{page:commentT1}When the number of groups is over-specified ($G \geq G_0,; H \geq H_0$), the parameter estimates remain consistent. The main reason is that, although some of the true underlying groups may be further partitioned into several subgroups, nodes within each subgroup still share the same group-level parameters. Consequently, our model assumptions remain valid, and the consistency of the estimators is unaffected by over-specification.
		By  \eqref{eq:d_w_conv},
the estimation error across all quantile levels
		 enjoys an $O_p(\ol{n}\log(NK)/T)$ convergence rate.
		This indicates that for a network with a large average network degree $\ol{n}$, the convergence becomes slower due to the potentially complex dependence structure.
 When $H = 1$, the rate is independent of $\ol{n}$ (see \cite{ren2026multi}),  but this does not extend to $H > 1$. See Section~G.1 of the supplementary material for details and sensitivity analysis.

		Based on \eqref{eq:d_w_conv}, we can proceed to obtain error bounds for the mis-clustering rates for both types of groups. To this end, define $\wh {\mathcal{R}}_g = \{i: \wh {g}_i = g\}$ and $\wh {\mathcal{C}}_h = \{j: \wh {h}_j = h\}$ as the set of estimated group members. To properly measure the clustering error, following \cite{zhu2025simultaneous}, we first define the following maps $\chi_1:[G]\to [G_0]$ and $\chi_2:[H]\to [H_0]$ as	
		\compressmath	
		\begin{align}
			&\chi_1(g) = \arg\max_{g'\in [G_0]} \sum_i I(i\in \wh\cR_g, g_i^0 = g'), g\in [G],\label{eq:chi_1}\\
			&\chi_2(h) = \arg\max_{h'\in [H_0]} \sum_j I(j\in \wh\mC_h, h_j^0 = h'), h\in [H].\label{eq:chi_2}
		\end{align}
		Here $\chi_1(g)$ and $\chi_2(h)$ map $g$ and $h$ to the true groups where most nodes in $\wh \cR_g$ and $\wh \mC_h$ belong.
		Based on this mapping,
		the membership estimation errors can be defined as
		\begin{align}\label{def: rho1}
			&\wh \varrho_{1} = \frac{1}{N}\sum_{g=1}^{G}\sum_{i=1}^{N}I\left(i\in\wh\cR_g, g_i^0\neq \chi_1(g)\right),
			&\wh \varrho_{2} = \frac{1}{N}\sum_{h=1}^{H}\sum_{j=1}^{N}I\left(j\in\wh\mC_h, h_j^0\neq \chi_2(h)\right).
		\end{align}
		We remark that $1-\wh \varrho_1$ and $1-\wh \varrho_2$  are widely used in the machine learning literature and are commonly referred to as the clustering purity \citep{schutze2008introduction}.
		The error bounds for $\wh \varrho_1$ and $\wh \varrho_2$ are established in the next theorem.
		
		\bet\label{thm:error_rate}
		Assume Assumptions \ref{assum:f_bound}--\ref{assum:sample_size},   if $G\ge G_0$ and $H\ge H_0$, then we have $\wh \vr_{1} =  O_p({\ol n \log(NK)}T^{-1} c_{G,NT}^{-1})$ and $\wh \vr_2 = O_p({\ol n \log(NK)}T^{-1}c_{G,NT}^{-1}c_{H,NT}^{-1}c_{w,NT}^{-1}+N^{-1}|\mM_H^c|)$, where $\mM_H$ is given in Assumption \ref{assum:net}.
		\eet

\label{page:commentT2}The proof of Theorem \ref{thm:error_rate} is in Section C.3 of the supplementary material.
The error bounds for the mis-clustering rates are derived using the convergence result for the parameters in Theorem \ref{thm:consistency}.
First,
the mis-clustering rate for the row group is $O_p({\ol n \log(NK)}T^{-1} c_{G, NT}^{-1})$, suggesting that a larger $T$ yields better clustering accuracy. This result is similar to those in \cite{liu2020identification,su2016identifying,zhang2019quantile, zhu2025simultaneous}.

However, the derivation of the clustering error bound for the column memberships is much more challenging and significantly different from the proofs in the existing literature.\label{page:remark-T2}
That is mainly because the row and column memberships play different roles as we comment in Section A of the supplementary material.
In addition, since both row and column memberships are involved in the loss function, they cannot be analyzed separately and should be investigated simultaneously.
In particular, the proof technique for establishing the error bound of $
\wh \vr_1$  cannot be directly borrowed, and new proof techniques are thus developed for analyzing $\wh \vr_2$.
Specifically, we establish the error bound for $
\wh \vr_1$ by using $\{\wh \nu_g(\tau_k), \wh \bgamma_g(\tau_k)\}$ with similar proof technique in \cite{zhu2025simultaneous}.
However, establishing the error rate for $\wh \vr_2$ totally relies on the analysis of $\{\wh \theta_{gh}(\tau_k)\}$, thus $c_{G,NT}^{-1}$ in Assumption \ref{assum:group_sep} occurs in the estimation error here.

As discussed in Section~A of the supplementary material,
the roles of $\mG$ and $\mH$ are asymmetric, and $\mH$
can be treated as a group structure in the ``feature dimension'',
which is unsolved in the existing literature in panel data
with network dependencies.
In our proof, we relate the error rate to the average estimation accuracy in \eqref{eq:d_w_conv} with a new requirement for the network structure in  Assumption \ref{assum:net}, under the control of the row group estimation error.
Basically, it requires that the in-degree information $q_j$ for nodes in $\mM_H$ should be sufficient to provide signal strength to reliably identify their influence groups.
As a consequence, the estimation error for the column group relates to both $c_{G,NT}$, $c_{H,NT}$, and $c_{w,NT}$, yielding a higher estimation error.
This phenomenon is also observed in our simulation studies, where the error rate of the column memberships is typically higher than the row memberships.\label{page:commentT2-end}

\vspace{-1.1em}
\subsection{Group Number Identification}\label{sec:selectG}
\vspace{-0.5em}

Note that the group numbers $(G,H)$ are pre-specified when optimizing the objective function \eqref{eq:loss_func}. The mis-specification of the group numbers may lead to unsatisfactory performance of the estimators. Therefore, it is crucial to develop a group number estimation procedure that can consistently estimate the true group numbers $(G_0,H_0)$. For pre-specified group numbers $(G,H)$, denote the corresponding estimators as $\wh \bpsi^{(G, H)}, \wh \mG^{(G, H)}, \wh \mH^{(G, H)}$ in this subsection by slightly abusing the notation. To select the number of row and column groups, we introduce the following quantile loss-based information criterion (QIC)
\compressmath
\begin{align}\label{eq:QIC}
   \qic_{\lambda_{NT}}(G, H) = \log\{\mL(\wh \bpsi^{(G, H)},
  \wh\mG^{(G, H)}, \wh\mH^{(G, H)})\} + \lambda_{NT}G(H+p+1),
\end{align}
where $\lambda_{NT}$ is a tuning parameter, and $G(H+p+1)$ is the number of model parameters for any given quantile. We can estimate $G$ and $H$ as $(\wh{G}, \wh{H}) = \arg\min_{G,H}\qic_{\lambda_{NT}}(G, H)$. In the following, we show that the group number can be consistently estimated when $\lambda_{NT}$ is appropriately chosen.
\vspace{-0.5em}
\bet\label{thm:number}
Suppose that Assumptions~\ref{assum:f_bound}--\ref{assum:sample_size} are satisfied. If
$\min (c_{G, N T}, c_{H, N T})\gg\lambda_{NT}$, and ${\lambda_{NT}T}/\{{\ol n \log(NK)}\}\to \infty$,
then we have that $P(\wh G = G_0, \wh H = H_0)\to 1$ as $(N,T)\to\infty$.
\eet
\vspace{-0.5em}

The proof of Theorem \ref{thm:number} is provided in Section C.4 of the supplementary material. The tuning parameter $\lambda_{NT}$ should
be of an appropriate order
 to ensure that the model is neither under-fitted nor over-fitted.
\label{page:commentT3}We note that the loss function decreases monotonically as $G, H$ increase, and QIC balances this through the penalty term $\lambda_{NT} G(H+p+1)$.
Under underfitting, true groups are forcibly merged, causing the loss to increase at the order of $\min(c_{G,NT}, c_{H,NT})$; the condition $\min(c_{G,NT}, c_{H,NT}) \gg \lambda_{NT}$ ensures this increase dominates the penalty.
Under overfitting, the loss decreases in the order of $O_p(\bar n \log(NK)/T)$, thus $\lambda_{NT} T/\{\ol n \log(NK)\} \to \infty$ ensures the penalty dominates this reduction.
This leads to the consistent choice of the true number of groups.
Since $K$ enters this condition only through $\log(NK)$, as long as $K$ increases at a polynomial rate in $N$, the term $\log(NK)$ remains of the same order as $\log(N)$. Therefore, in our empirical specification we use $\lambda_{NT}=N^{1 / 10} T^{-1}\log(T)/(10\min(\ol{n}, 10))$, since $N^{1/10}$ grows substantially faster than $\log(NK)$ as $N$ grows in general. Our empirical findings also suggest that excluding $K$ from $\lambda_{NT}$ leads to more stable empirical performance.
Robustness of the QIC criterion is further examined in Section G.2 of the supplementary material.

Additionally, we note that establishing the consistency result for $\wh{H}$ is more challenging than for $\wh {G}$.
This is primarily due to the fact that the clustering error for the $\mG$ memberships is inseparable from the theoretical analysis of $\wh{H}$ and requires careful attention.
\label{page: r_thm3}In particular, we remark that the critical challenge is to establish the lower bound
for $(NK)^{-1}\sum_{i=1}^N\sum_{k=1}^K\sum_{j} w_{ij}^2
|\wh\theta_{\wh g_i\wh h_j}(\tau_k) - \theta_{g_i^0h_j^0}^0(\tau_k)|^2$ when $H<H_0$, which is not discussed in existing literature since the two-way memberships and the network dependences are involved.
In our proof, we relate this lower bound to two important quantities.
The first is the true group differences given by $\min_{h_0\ne h_0'\in[H_0],g_0\in[G_0]}	(K^{-1}\sum_{k=1}^{K}|\theta_{g_0h_0}^0(\tau_k) - \theta_{g_0h_0'}^0(\tau_k)|^2)$ in Assumption \ref{assum:group_sep}, which reflects the signal strength to distinguish different column groups.
The second is the
 $\pi_{g_0h_0}$ given in Assumption \ref{assum:c_pi}, which characterizes the effective cross-group
links.
These two quantities jointly determine the identifiability of the group structure when the number of groups is underestimated,
and they serve as the foundation for the consistent results established in the subsequent theoretical analysis.

\vspace{-1.1em}
\section{Model Inference When $G=G_0$ and $H=H_0$}\label{section: inference}
\vspace{-0.5em}

In this section, we discuss the model inference based on the estimated TGNQ model under the assumption that $G=G_0$ and $H=H_0$. To establish a valid inference procedure, we first conduct membership refinements for $\wh \mG$ and $\wh \mH$, respectively, which are necessary to establish asymptotic normality of model parameter estimators.

\vspace{-1.1em}
\subsection{Membership Refinement}
\vspace{-0.5em}
\subsubsection{Refinement of $\mG$ Membership}
\vspace{-0.5em}

Let $\mH_i =(h_j: j \in \mN_i)^\top  \in \mR^{n_i}$ denote the group memberships of
nodes followed by the $i$th node.
Let $\btheta_{g_i, \mH_i}(\tau_k) =$
$(\theta_{g_ih_j}(\tau_k): j\in \mN_i)^\top \in \mR^{n_i}$, $\bzeta_{g_i}(\tau_k) = (\nu_{g_i}(\tau_k), \bgamma_{g_i}^\top(\tau_k) )^\top \in \mR^{p+1}$
and correspondingly, the concatenated parameter vectors as
$\btheta_{g_i, \mH_i}=(\btheta_{g_i, \mH_i}^\top (\tau_1),\cdots,\btheta_{g_i, \mH_i}^\top (\tau_K))^\top\in \mR^{n_iK}$ and $\bzeta_{g_i}=(\bzeta_{g_i}^\top (\tau_1),\cdots,\bzeta_{g_i}^\top (\tau_K))^\top\in \mR^{K(p+1)}$, for
$i \in [N]$.
For convenience, we rewrite the loss function of the node $i$ as
$\mL_i(\btheta_{g_i,\mH_i}, \bzeta_{g_i})={(KT)^{-1}}\sum_{k=1}^{K}\sum_{t=1}^T
\rho_{\tau_k} (Y_{it} - \sum_j\theta_{g_i h_j}(\tau_k)w_{ij}Y_{j(t-1)} - \nu_{g_i}(\tau_k) Y_{i(t-1)}- \bx_{it}^\top\bgamma_{g_i}(\tau_k))$
in this subsection.
Specifically, denote
$\wh \bPhi_i =\{\wh\btheta_{g\mH}: g \in [G], \mH \in [H]^{n_i}\}$.
As a result, $\wh \bPhi_i$ collects possible estimated network interaction effects at $K$ quantiles by exhausting all group memberships.
Then we define the following profiled node-specific loss function as
\compressmath
\begin{align*}
\mL_i^P(g) = \min_{\bvarphi_i\in \wh \bPhi_i}\mL_{i}(\bvarphi_i,
\wh \bzeta_{g}).
\end{align*}
Define $\wh g_i^\dag = \arg\min_{g\in [G]} \mL_i^P(g)$.
If $\mL_i^P(\wh g_i^\dag)$ is sufficiently smaller than
$\mL_i(\wh\btheta_{\wh g_i,\wh\mH_i}, \wh\bzeta_{\wh g_i})$,
we then refine the membership estimate from $\wh g_i$ to $\wh g_i^\dag$ using the following refinement protocol
\compressmath
\beq
\wh g_i^r=
				\begin{cases}
					\wh g_i, & \mbox{if }  \mL_i(\wh\btheta_{\wh g_i,\wh\mH_i}, \wh\bzeta_{\wh g_i}) - \mL_i^P({\wh g_i^\dag})\le \frac{1}{\sqrt{T}} \mL_i^P({\wh g_i^\dag}),\\
					\wh g_i^\dag, & \mbox{if }
					  \mL_i(\wh\btheta_{\wh g_i,\wh\mH_i}, \wh\bzeta_{\wh g_i}) -  \mL_i^P({\wh g_i^\dag}) > \frac{1}{\sqrt{T}} \mL_i^P({\wh g_i^\dag}).\nonumber
				\end{cases}
\eeq
\vspace{-1.5em}
\subsubsection{Refinement of $\mH$ Membership}
\vspace{-0.5em}
Next, we refine the $\mH$ membership of the $j$th node.
Recall that for the $j$th node, $\mathbb{F}_j = \{i: a_{ij} = 1\}$ is the set of its followers.
The change of the membership $h_j$ may affect the memberships in $\mH_i$ for $i\in \mathbb{F}_j $.
Denote the collection of nodes in $\mN_i$ for $i\in \mathbb{F}_j $ as
$\mathbb{F}_j ^2 = \{l: \sum_i a_{il}a_{ij}\ne 0,l\ne j\}$, and additionally define $\wt \mH_j = (h_l: l\in \mathbb{F}_j ^2)^\top$.
As a result, $\mathbb{F}_j ^2$ collects the nodes that have a second-order relationship with node $j$, whose column memberships can be affected by the change of $h_j$.
Correspondingly, $\wt\mH_j$ collects their {column} memberships and let $ o_j = |\mathbb{F}_j ^2|$.
We define the following loss function for node $j$ by fixing $h_j = h$ and exhausting possible memberships in $\wt\mH_j$,
\compressmath
\begin{align*}
\cH_j^P(h)=\min_{\widetilde{\mH}_j\in[H]^{o_j},h_j=h}\sum_{i = 1}^N a_{ij}\mL_i( \wh \btheta_{\wh g_i^r,\check \mH_i },
\wh \bzeta_{\wh g_i^r}),\quad j=1,\cdots,N,
\end{align*}
where $\check \mH_i\in \mR^{n_i}$ is obtained by fixing $h_j = h$ and the other memberships are taken from $\wt \mH_j$,
In addition, the $\mG$ memberships are fixed as $g_i = \wh g_i^r$ for all $i\in[N]$, and the estimated model parameters are also fixed.

Define $\wh h_j^\dag = \arg\min_{h\in [H]} \cH_j^P(h)$.
If $\cH_j^P(\wh h_j^\dag)$ is sufficiently smaller than its unrefined counterpart
$\sum_{i = 1}^N a_{ij}\mL_i(\wh\btheta_{\wh g_i^r,\wh\mH_i}, \wh\bzeta_{\wh g_i^r})$,
we can refine the membership estimation $\wh h_j$ to $\wh h_j^\dag$.
Specifically, the following refinement protocol is used,
\compressmath
\beq
\wh h_j^r=
\begin{cases}
	\wh h_j, & \mbox{if }  \sum_{i = 1}^N a_{ij}\mL_i(\wh\btheta_{\wh g_i^r,\wh\mH_i}, \wh\bzeta_{\wh g_i^r}) - \cH_j^P(\wh h_j^\dag )\le \frac{1}{\sqrt{T}} \cH_j^P(\wh h_j^\dag )\\
	\wh h_j^\dag, & \mbox{if }
	\sum_{i = 1}^N a_{ij}\mL_i(\wh\btheta_{\wh g_i^r,\wh\mH_i}, \wh\bzeta_{\wh g_i^r}) - \cH_j^P(\wh h_j^\dag )> \frac{1}{\sqrt{T}} \cH_j^P(\wh h_j^\dag ).\label{eq:refine_H}
\end{cases}
\eeq
After obtaining the refined memberships $\wh \mG^r = (\wh g_i^r: i\in [N])^\top$ and $\wh \mH^r = (\wh h_j^r: j\in [N])$, one recomputes the post-refined estimator $\wh \bpsi^r = \arg\min_{\bpsi}\mL(\bpsi, \wh \mG^r, \wh \mH^r)$. Next, we show the strong consistency for $\wh \mG^r$ and $\wh \mH^r$, which leads to the asymptotic normality of $\wh \bpsi^r$.

\vspace{-0.5em}
\begin{remark}\label{remark:refine}
 The refinement procedure is used to establish the strong consistency result, which further implies the oracle property that the post-refinement estimator is asymptotically equivalent to the oracle estimator with known true memberships, thereby ensuring valid statistical inference.
 In addition, we would like to emphasize that this procedure is not required for the general consistency results in Theorems \ref{thm:consistency}-\ref{thm:number}.
Unlike the traditional GPQR model \citep{zhang2019quantile}, where independent individuals allow the total loss to decompose into individual-level losses, our network setting involves complicated interactions that prevent such decomposition, necessitating the refinement procedure.
        Therefore, we utilize a further refinement procedure to facilitate our theoretical analysis for tackling this issue.
     We also report the pre-refinement estimation result in our simulation study in Section B.5 of the supplementary material, and we find the estimation accuracies are quite comparable.

\end{remark}

\vspace{-1.1em}
\subsection{Theoretical Analysis}
\vspace{-0.5em}

\subsubsection{Membership Estimation Consistency}
\vspace{-0.5em}

We now show that all group memberships can be consistently estimated after membership refinements.
Let $\wh \cR_g^r = \{i: \wh g_i^r = g\}$, $\cR_{g_0}^0 = \{i: g_i^0 = g_0\}$,
$\wh \mC_h^r = \{j: \wh h_j^r = h\}$ and $\mC_{h_0}^0 = \{j: h_j^0 = h_0\}$.
We first strengthen Assumption \ref{assum:sample_size}(b) to hold uniformly over $i\in[N]$.
\vspace{-0.5em}
\begin{assumption}
		\label{assum:nmax}
			As $(N,T)\to \infty$, assume
$c_{G, NT} \gg \sqrt{{n_{\max}\log(NK)}/{T}}+{(\log(NKT))^{1/\gamma}}/{T} $.
\end{assumption}
\vspace{-0.5em}
Based on this, the next theorem establishes the $\mG$ membership estimation consistency.
\bet\label{thm:g_consist}
Under Assumptions \ref{assum:f_bound}--\ref{assum:nmax} with $G= G_0$ and $H=H_0$, we have \\
(a) $			\sup_{i\in[N]} {K}^{-1}\sum_{k=1}^{K}\|\wh \bzeta_{\wh g_i^r}(\tau_k) - \bzeta_{g_i^0}^0(\tau_k)\|^2 = o_p( 1)$;\\
(b) For any $g\in [G]$, there exists one $g_0\in [G_0]$ such that P($\wh \cR_g^r = \cR_{g_0}^{0})\to 1$.
\eet	
\vspace{-0.5em}
The proof of Theorem \ref{thm:g_consist} is provided in Section C.5 of the supplementary material.

Subsequently, we proceed to establish the $\mH$ membership estimation consistency, which further requires the following assumption.

\vspace{-0.5em}
\begin{assumption}\label{assum:w_ij}
  Assume that (a) $\sup_{j\in[N]}d_j^{-1}\min_{h\ne h_j^0}\{K^{-1}\sum_{k=1}^{K}\sum_{i=1}^Nw_{ij}^2
  |\theta_{g_i^0h_j^0}^0(\tau_k) - \theta_{g_i^0h}^0(\tau_k)|^2
  \}\ge c_0$ with $d_j=\sum_{i=1}^N a_{ij}$ being the in-degree of node $j$; (b)
  \compressmath
\beq
\max_j q_j\Big(\frac{\ol n^2 \log(NK)}{T c_{G,NT}c_{H,NT}{c_{w,NT}}} + N^{-1}\ol n |\mM_H^c|\Big)+\frac{\ol n^2 \log(NK)}{Tc_{G,NT}}= o(1);\label{eq:w_ij_sum}
\eeq
and that (c) $N^{-1}\sum_j q_j^2 \le c$, where
			$q_j = \sum_i w_{ij}^2$ and $c$ is a positive constant.
\end{assumption}
\vspace{-0.5em}
Assumption \ref{assum:w_ij} (a) imposes a slightly more restrictive condition on interaction parameters concerning $\mH$ group differences than Assumptions \ref{assum:group_sep}. 
 Assumption \ref{assum:w_ij} (b) is related to the nodes' in-degrees, which puts restrictions on the diverging speed of $\max_j q_j$. 
Under this condition, we can further prove the following Theorem.

\bet\label{thm:h_consist}
Under  Assumptions \ref{assum:f_bound}--\ref{assum:w_ij}, if $G= G_0$ and $H = H_0$, then  \\
(a) $\sup_{j\in[N]}\{{(d_jK)^{-1}}\sum_{i = 1}^N \sum_{k = 1}^Kw_{ij}^2|\wh\theta_{\wh g_i^r \wh h_j^r}(\tau_k)-\theta_{g_i^0h_j^0}^0(\tau_k)|^2\} = o_p(1)$;\\
(b) For any $h\in [H]$, there exists one $h_0\in [H_0]$ such that P($\wh \mC_h^r = \mC_{h_0}^{0})\to 1$.
\eet	

The proof of Theorem \ref{thm:h_consist} is given in Section C.6 of the supplementary material.  The consistency for $\mH$ memberships is relatively more difficult to show than for the $\mG$ memberships. Specifically, in conclusion (a), we first establish the uniform estimation consistency for $\wh \theta_{gh}(\tau_k)$ using the refined memberships ${\wh h_j^r}$. This enables us to further obtain the estimation consistency for $\wh h_j^r$ as stated in conclusion (b). Lastly, with the result in Theorems \ref{thm:g_consist} and \ref{thm:h_consist}, we can conclude that the post-refined estimator $\wh \bpsi^r$ is asymptotically equivalent to the oracle estimator as if the true memberships were known in advance. This enables us to conduct valid statistical inferences.
We can further relax the lower bound $c_0$ in Assumption \ref{assum:w_ij} as a sequence
$\wt c_{H,NT}\to 0$, and we include the corresponding results in Section F of the supplementary material.

\vspace{-1.1em}
\subsubsection{Asymptotic Normality}
\vspace{-0.5em}

Note that when fixing $\mG = \wh\mG^r$ and $\mH = \wh\mH^r$, minimizing \eqref{eq:loss_func} is equivalent to minimizing
$\mL_{\tau_k}(\bpsi,\wh \mG^r,\wh \mH^r)$ separately at each $\tau_k$.
Let
${\bxi( \tau_k)} = (\bxi_1(\tau_k)^\top,\cdots, \bxi_{G_0}(\tau_k)^\top)^\top \in \mR^{G_0(H_0+p+1)}$ with $\bxi_g(\tau_k)$
defined in~\eqref{eq:d}, and write $\cX_{it} = \e_{g_i^0}^{(G_0)}\otimes (\ol \y_i^{\top},Y_{i(t-1)},\x_{it}^\top)^\top\in \mR^{G_0(H_0+p+1)}$ with $\ol \y_i  = (\ol Y_h: h \in[H_0])^\top\in  \mR^{H_0}$ and $\ol Y_h = \sum_j I(h_j^0 = h)w_{ij}Y_{j(t-1)}$. Then we can express the model~\eqref{gbnar} as
\compressmath
\begin{align*}
\y_t = \mX_t \bxi(\tau_k) + \bve_t(\tau_k),
\end{align*}
where $\mX_t = (\cX_{it}: i\in [N])\in \mR^{N\times G_0(H_0+p+1)}$
and $\bve_t(\tau_k) = (\ve_{it}(\tau_k): i\in [N])^\top\in \mR^N$.
 Denote the true parameter as $\bxi^0 = (\bxi^0(\tau_1)^\top,\cdots,\bxi^0(\tau_K)^\top)^\top$, and
 define $\wh \bxi^r(\tau_k)$ as the post-refined estimator when substituting $\mG = \wh \mG^r$ and $\mH = \wh \mH^r$.
To establish the asymptotic normality of $\wh \bxi^r(\tau_k)$, we further require the following conditions.

\vspace{-0.5em}
\begin{assumption}\label{assum:Sig_cX}
Assume that
(a) $\bSigma_f(\tau_k) = \lim_{N\to\infty}N^{-1}\sum_{i=1}^N \E\{f_{i\tau_k}(0|\cX_{it})\cX_{it}\cX_{it}^\top\}$ exists and nonsingular;
(b) denote $\g_t(\tau_k) = N^{-1/2}\sum_{i = 1}^N\{\tau_k - I(\ve_{it}(\tau_k)<0)\}\cX_{it}\in \mR^{G_0(H_0+p+1)}$ and
 assume
$\bSigma_\cX(\tau_k) = \lim_{N,T\to\infty} \cov(T^{-1/2}\sum_{t = 1}^T\g_t(\tau_k))$ exists and is nonsingular.
Let $\E(|\bfeta^\top\g_t(\tau_k)|^\delta)<\infty$ for some constant $\delta>2$ with any vector $\bfeta$ satisfying $\|\bfeta\| = 1$; and (c) $\gamma_1\ge 1$ in Assumption \ref{assum:beta_mixing}.
\end{assumption}
\vspace{-0.5em}
 Assumption~\ref{assum:Sig_cX} (a) and (b) specify a set of moment conditions related to $\mX_t$, where the matrices $\bSigma_f(\tau_k)$ and $\bSigma_\cX(\tau_k)$ play important roles in the asymptotic covariance of $\wh \bxi^r(\tau_k)$ as stated in Theorem \ref{thm:normal}. Assumption~\ref{assum:Sig_cX} (c) prevents the tail of the distribution of $Y_{it}$ from being too heavy. Compared to existing literature \citep{kato2012asymptotics,zhang2019quantile}, which typically assumes a bounded covariates assumption, condition (c) is more relaxed.

\bet\label{thm:normal}
Assume that Assumptions \ref{assum:f_bound}--\ref{assum:Sig_cX} hold.
Then we have that
\compressmath
\begin{align*}
\sqrt{NT}(\wh \bxi^r(\tau_k) - \bxi^0(\tau_k))\to_d N\left(\zero,  \bSigma_f(\tau_k)^{-1}\bSigma_\cX(\tau_k)\bSigma_f(\tau_k)^{-1}\right).
\end{align*}
\eet

The proof of Theorem \ref{thm:normal} is provided in Section C.7 of the supplementary material.
The proof follows the theoretical framework developed by \cite{kato2012asymptotics} for quantile panel data models, and the central limit theorems for dependent time series in
\cite{fan2003nonlinear}.
We remove the restriction that $\cX_{it}$ is uniformly bounded as required in \cite{kato2012asymptotics} to adapt for the applications in our scenario.

\vspace{-1.1em}
\section{Simulation Studies}\label{section: simu}
\vspace{-0.5em}

To examine the finite-sample performance of the proposed method, we conducted extensive simulation studies.
Following \cite{koenker2006quantile} and \cite{zhu2019network},
we generate the data using the following data-generating process
\compressmath
\beq
Y_{it} = \sum_{j = 1}^N{\theta_{g_ih_j}(U_{it})}w_{ij}Y_{j(t-1)} + \nu_{g_i}(U_{it}) Y_{i(t-1)} + \mathbf{x}_{it}^\top\bgamma_{g_i}(U_{it}),\label{eq:simu_model}
\eeq
where recall that $w_{ij} = a_{ij}/n_i$ denotes the element of the row-normalized weighting matrix.
In this context, $U_{it}$'s are independently distributed variables following a uniform distribution between 0 and 1,
which acts as noise in the data generation process.
Additionally, $\theta_{g_ih_j}(\cdot)$, $\nu_{g_i}(\cdot)$ and $\bgamma_{g_i}(\cdot)$  are non-decreasing functions governing parameters of different groups.
Assuming the right side of equation \eqref{eq:simu_model} is an increasing function of $U_{it}$,  the conditional quantile function $Q_{it}(\tau|  \bx_{it}, \mF_{t-1})$  consistent with \eqref{gbnar} can be written as
\compressmath
\begin{align*}
	Q_{it}(\tau|  \bx_{it}, \mF_{t-1}) = \sum_{j = 1}^N\theta_{g_i h_j}(\tau)w_{ij}Y_{j(t-1)} + \nu_{g_i}(\tau) Y_{i(t-1)} + \bx_{it}^\top\bgamma_{g_i}(\tau).
\end{align*}

Across all settings, the covariate vectors $\bx_{it}$'s $\in \mR^p$ are independently generated {as absolute values of} a multivariate normal distribution $N(\mathbf{0}, \bI_p)$ with $p=2$.
For each network structure, we consider two settings with $G_0=H_0=2$ and $G_0=H_0=3$.
When $G_0 = H_0 = 2$, the row membership ratios are determined by the parameters $(\pi_1^G, \pi_2^G) = (0.5, 0.5)$, and the column membership ratios are determined by $(\pi_1^H, \pi_2^H) = (0.4, 0.6)$.
In the case of $G_0 = H_0 = 3$, we set the row and column membership ratios to be $(\pi^G_1, \pi^G_2, \pi^G_3)=(\pi^H_1, \pi^H_2, \pi^H_3) = (0.3, 0.3, 0.4)$.
We consider two distinct network structures: the Stochastic Block Model (SBM) and Power-Law Distribution Network \citep{clauset2009power}, with detailed generation procedures provided in Section B.4.1 of the supplementary material.

Under each network structure, we evaluate the performance of the proposed method under two different parameter settings.
In \textsc{Scenario} 1, we generate data using an additive model,
specifying $\theta_{g_ih_j}(\tau)$ in an additive form as $\theta_{g_ih_j}(\tau) =\alpha_{g_i}(\tau)+\beta_{h_j}(\tau)$,
while in \textsc{Scenario} 2, we generate data using a multiplicative model, where the network effects are considered as $\theta_{g_ih_j}(\tau) =\alpha_{g_i}(\tau)\times\beta_{h_j}(\tau)$.
The details of these functions are provided in Section B.4.1 of the supplementary material.

\vspace{-1.1em}
\subsection{Estimation and Inference when $G=G_0$ and $H=H_0$}\label{section: simulation-correct}
\vspace{-0.5em}

First, we evaluate the estimation performances when $G=G_0$ and $H=H_0$.
We apply the general estimation procedure stated in Section \ref{sec:model_est} for both scenarios.
Furthermore, we exploit the special model structures (i.e., additive model and multiplicative model) for estimation in {\sc Scenario 1} and {\sc Scenario 2} respectively with estimation procedures given in Section B.3 of the supplementary material,
where the estimation results can be found in Section B.4.2 of the supplementary material to save space here.
In both scenarios, we also evaluate the oracle estimators when the true memberships are given for comparison.
We conduct model estimation separately at five quantile levels, i.e., ($\tau_1, \tau_2, \tau_3, \tau_4, \tau_5) = (0.1, 0.3, 0.5, 0.7, 0.9)$ for $B = 500$ repeated experiments.
At each quantile level, the following metrics are calculated for measuring the estimation accuracy.
For brevity, we omit the quantile index $\tau_k$ of the parameters in the subsequent sections.
Let $\wh {\btheta}^{r(b)}$, $\wh {\bnu}^{r(b)}$ and $\wh {\bgamma}^{r(b)}$ be the refined estimators obtained from the $b$th simulation round.
We calculate the root mean squared error (RMSE) for $\btheta$ as
$\text{RMSE}_{\btheta}=B^{-1} \sum_{b=1}^B\| \wh {\btheta}^{r(b)}-\btheta^0\|$.
The RMSE for the other estimators can be similarly defined.
Next, we conduct statistical inference for each estimator and
construct 95\% confidence intervals (CI) for each model parameter.
Taking $\bnu^0$ as an example, in the $b$th simulation round, we construct the 95\% CI for $\nu_g^0$ as $\text{CI}_{\bnu,g}^{(b)}=(\wh {\nu}_g^{r(b)}-1.96 \wh {\text{SE}}_{\nu,g}^{(b)},\quad \wh {\nu}_g^{r(b)}+1.96 \wh {\text{SE}}_{\nu,g}^{(b)})$,
where $\wh {\text{SE}}_{\nu,g}^{(b)}$ is the estimated asymptotic standard error of $\wh {\nu}_g^{r(b)}$ based on Theorem \ref{thm:normal}.
For all components of $\bnu$, the average coverage error is computed as $\text{AE}_{\text{cp}, \bnu}=G_0^{-1} \sum_{g=1}^{G_0}|B^{-1} \sum_{b=1}^B I(\nu_g^0 \in \text{CI}_{\nu,g}^{(b)})-0.95|$.
Lastly, we calculate the row group membership estimation error as $\wh \varrho_1^r=(NB)^{-1} \sum_{b=1}^B\sum_{i=1}^N I(\wh g_i^{r(b)}\neq g_i^0)$, where $\wh g_i^{r(b)}$ corresponds to the refined row group membership estimator for node $i $ from the $b$th simulation round (after label permutation).\label{page: sim_rho_1_def}
The column group membership estimation error rate  $\wh \varrho_2^r$ is defined similarly.

The simulation results in {\sc Scenario 1} for the SBM network are summarized in Tables \ref{table: scenario-1-ge}.
Simulations in {\sc Scenario 2} for the SBM network and those conducted on the power-law networks yield similar results, details of which can be found in Section B.4.2 of the supplementary material.
First, Table \ref{table: scenario-1-ge}
shows that the accuracy of parameter estimation steadily improves
as $N$ or $T$ increases at all quantile levels.
Similarly, the group membership estimation error rates $\wh \varrho^r_1$ and $\wh \varrho^r_2$ significantly decrease as $N$ and $T$ grow.
As expected, for fixed $N$ and $T$, the proposed method performs much better under $G_0=2$ than under $G_0=3$, because the former has more data within each group to yield better estimation results.
Furthermore, we observe lower $\text{AE}_{\text{cp}}$ values for both scenarios when $N$ and $T$ are larger, which indicates a valid inference result.
As $N$ and $T$ increase, the performance of the proposed estimation gradually approaches that of the oracle estimation.
This lends further support to our theoretical findings in Theorem \ref{thm:normal}.

\begin{sidewaystable}
\centering
\caption{RMSEs ($\times 10^{-2}$) and $\text{AE}_{\text{cp}}$'s  (\%, in parentheses) in \textsc{Scenario} 1 for the SBM network.}\label{table: scenario-1-ge}
\setlength{\tabcolsep}{1 pt}
{\fontsize{6.5}{7}\selectfont
\begin{tabular*}{\textwidth}{@{\extracolsep{\fill}}ccrrrrrrrrr ccrrrrrrrrr@{\extracolsep{\fill}}}
\toprule
\multicolumn{11}{c}{$G=2$}&\multicolumn{11}{c}{$G=3$}\\
\cline{1-11}\cline{12-22}
&&&\multicolumn{3}{c}{Oracle estimator}&\multicolumn{3}{c}{TGNQ with refinement}&&&&&&\multicolumn{3}{c}{Oracle estimator}&\multicolumn{3}{c}{TGNQ with refinement}\\
\cline{4-6}\cline{7-11} \cline{15-17}\cline{18-22}\\
$N$&$T$&$\tau$&$\wh \btheta_o$&$\wh \bnu_o$&$\wh \bgamma_o$&$\wh \btheta^r$&$\wh \bnu^r$&$\wh \bgamma^r$&$\wh \varrho_1^r(\%)$&$\wh \varrho_2^r(\%)$&$N$&$T$&$\tau$&$\wh \btheta_o$&$\wh \bnu_o$&$\wh \bgamma_o$&$\wh \btheta^r$&$\wh \bnu^r$&$\wh \bgamma^r$&$\wh \varrho_1^r(\%)$&$\wh \varrho_2^r(\%)$\\
\cline{1-11}\cline{12-22}
 100 & 50 & 0.1 & 5.0 (3.8) & 1.8 (3.6) & 0.9 (1.1) & 6.1 (11.3) & 1.9 (4.2) & 1.0 (2.5) & 2.8 & 6.8 & 100 & 100 & 0.1 & 7.6 (2.8) & 2.0 (2.7) & 1.0 (1.4) & 14.4 (24.4) & 2.2 (5.1) & 1.1 (4.8) & 9.7 & 19.2\\
     &  & 0.3 & 6.4 (1.7) & 2.4 (2.0) & 1.2 (1.2) & 7.6 (8.4) & 2.5 (2.6) & 1.3 (3.4) &  &  &  &  & 0.3& 10.2 (1.6) & 2.7 (1.2) & 1.3 (1.2) & 17.9 (20.4) & 3.0 (4.3) & 1.5 (5.5) &  &  \\
     &  & 0.5 & 6.4 (1.2) & 2.5 (1.6) & 1.2 (1.4) & 7.8 (8.7) & 2.6 (2.0) & 1.3 (2.7) &  &   &  &  & 0.5 & 10.7 (1.2) & 3.0 (0.9) & 1.4 (0.9) & 19.2 (21.7) & 3.3 (4.3) & 1.7 (5.7) &  &  \\
     &  & 0.7 & 5.6 (2.6) & 2.2 (1.0) & 1.0 (1.0) & 6.8 (9.7) & 2.2 (1.0) & 1.1 (2.6) &  &  &  &  & 0.7 & 9.7 (1.7) & 2.6 (1.5) & 1.3 (1.7) & 16.9 (20.6) & 2.9 (4.3) & 1.4 (5.9) &  &  \\
     &  & 0.9 & 3.7 (3.2) & 1.4 (1.3) & 0.7 (1.1) & 5.1 (12.8) & 1.6 (2.2) & 0.7 (2.5) &  & &  &  & 0.9 & 7.0 (3.9) & 1.9 (2.5) & 0.9 (0.7) & 13.7 (28.0) & 2.1 (3.7) & 1.0 (5.6) &  &  \\ \cline{1-11}\cline{12-22}
    100 & 100 & 0.1 & 3.5 (3.2) & 1.2 (1.4) & 0.7 (0.7) & 3.5 (4.2) & 1.2 (1.6) & 0.7 (1.0) & 0.1 & 0.5 & 100 & 200 & 0.1 & 5.4 (2.3) & 1.4 (1.4) & 0.7 (1.0) & 7.0 (10.6) & 1.4 (2.0) & 0.7 (1.8) & 3.0 & 7.6 \\
     &  & 0.3 & 4.5 (1.7) & 1.7 (1.0) & 0.8 (1.4) & 4.5 (2.1) & 1.7 (0.8) & 0.8 (1.2) &  &  &  &  & 0.3 & 7.2 (1.2) & 1.9 (1.1) & 0.9 (1.4) & 8.9 (7.5) & 2.0 (1.7) & 0.9 (1.8) &  &  \\
     &  & 0.5 & 4.5 (1.4) & 1.7 (0.4) & 0.8 (1.7) & 4.6 (1.8) & 1.7 (0.4) & 0.8 (1.9) &  &  &  &  & 0.5 & 7.5 (1.0) & 2.1 (0.9) & 1.0 (1.3) & 9.4 (7.1) & 2.2 (1.3) & 1.0 (1.6) &  &  \\
     &  & 0.7 & 3.9 (1.8) & 1.5 (1.0) & 0.7 (0.5) & 4.0 (2.5) & 1.5 (1.1) & 0.7 (0.8) &  &  &  &  & 0.7 & 6.9 (1.6) & 1.9 (0.9) & 0.9 (0.8) & 8.5 (8.3) & 2.0 (1.9) & 0.9 (1.6) &  &  \\
     &  & 0.9 & 2.6 (1.8) & 1.0 (0.8) & 0.5 (0.7) & 2.6 (2.7) & 1.0 (0.9) & 0.5 (1.0) &  &  &  &  & 0.9 & 4.9 (2.4) & 1.3 (0.5) & 0.6 (1.1) & 6.8 (14.5) & 1.4 (1.5) & 0.6 (1.6) &  &  \\ \cline{1-11}\cline{12-22}
    100 & 200 & 0.1 & 2.5 (1.6) & 0.9 (1.7) & 0.5 (0.7) & 2.5 (1.6) & 0.9 (1.9) & 0.5 (0.8) & 0.0 & 0.0 & 200 & 200 & 0.1 & 3.8 (2.2) & 1.0 (1.0) & 0.5 (0.9) & 4.9 (9.7) & 1.0 (1.7) & 0.5 (2.5) & 2.8 & 5.9 \\
     &  & 0.3 & 3.2 (0.8) & 1.2 (0.9) & 0.6 (0.7) & 3.2 (0.9) & 1.2 (1.0) & 0.6 (0.7) &  &  &  &  & 0.3 & 4.8 (1.1) & 1.4 (0.8) & 0.7 (0.4) & 5.8 (6.0) & 1.4 (1.4) & 0.7 (1.7) &  &  \\
     &  & 0.5 & 3.2 (1.2) & 1.3 (1.4) & 0.6 (0.5) & 3.2 (1.2) & 1.3 (1.4) & 0.6 (0.5) &  &  &  &  & 0.5 & 5.0 (0.8) & 1.5 (0.8) & 0.7 (0.5) & 5.9 (5.8) & 1.5 (1.8) & 0.7 (1.6) &  &  \\
     &  & 0.7 & 2.8 (1.9) & 1.1 (0.2) & 0.5 (0.8) & 2.8 (1.9) & 1.1 (0.3) & 0.5 (0.9) &  &  &  &  & 0.7 & 4.4 (0.7) & 1.3 (1.6) & 0.6 (1.0) & 5.3 (5.6) & 1.4 (1.8) & 0.6 (2.1) &  &  \\
     &  & 0.9 & 1.9 (2.2) & 0.7 (0.1) & 0.3 (0.3) & 1.9 (2.3) & 0.7 (0.1) & 0.3 (0.3) &  &  &  &  & 0.9 & 3.2 (1.9) & 1.0 (1.3) & 0.4 (0.5) & 4.3 (11.1) & 1.0 (2.1) & 0.4 (1.3) &  &  \\ \cline{1-11}\cline{12-22}
    200 & 200 & 0.1 & 1.6 (1.4) & 0.6 (1.2) & 0.3 (0.7) & 1.6 (1.5) & 0.6 (1.3) & 0.3 (0.8) & 0.0 & 0.0  & 200 & 400 & 0.10 & 2.7 (1.7) & 0.7 (1.1) & 0.4 (1.1) & 2.8 (3.2) & 0.7 (0.9) & 0.4 (1.3) & 0.3 & 1.4 \\
     &  & 0.3 & 2.1 (1.1) & 0.9 (0.7) & 0.4 (0.6) & 2.1 (1.1) & 0.9 (0.7) & 0.4 (0.7) &  &  &  &  & 0.30 & 3.4 (1.0) & 1.0 (0.7) & 0.5 (0.6) & 3.5 (1.7) & 1.0 (0.9) & 0.5 (0.6) &  &  \\
     &  & 0.5 & 2.1 (0.6) & 0.9 (0.3) & 0.4 (0.4) & 2.1 (0.6) & 0.9 (0.4) & 0.4 (0.5) &  &  &  &  & 0.50 & 3.5 (0.5) & 1.0 (0.7) & 0.5 (0.6) & 3.6 (1.0) & 1.0 (0.7) & 0.5 (0.4) &  &  \\
     &  & 0.7 & 1.7 (0.5) & 0.7 (1.0) & 0.3 (0.7) & 1.8 (0.5) & 0.7 (1.0) & 0.3 (0.7) &  &  &  &  & 0.70 & 3.1 (0.7) & 0.9 (0.8) & 0.4 (0.8) & 3.2 (1.2) & 0.9 (1.1) & 0.4 (0.5) &  &  \\
     &  & 0.9 & 1.2 (1.5) & 0.6 (0.9) & 0.2 (0.8) & 1.2 (1.6) & 0.6 (0.9) & 0.2 (0.8) &  &  &  &  & 0.90 & 2.3 (0.9) & 0.7 (0.3) & 0.3 (1.2) & 2.4 (3.6) & 0.7 (0.9) & 0.3 (1.3) &  &  \\
\Xhline{2\arrayrulewidth}
\end{tabular*}
}
\centering
\caption{RMSEs ($\times 10^{-2}$) in \textsc{Scenario} 1 for the SBM network and varying working models.}\label{table: mis-add}
{\fontsize{6}{6}\selectfont
\setlength{\tabcolsep}{1.5pt}
\begin{tabular*}{\textwidth}{@{\extracolsep{\fill}}cccrrrrrrrrrrrrrrrcccrrrrrrrrrrrrrrr@{\extracolsep{\fill}}}
\toprule
&&&\multicolumn{15}{c}{$G$=2}&\multicolumn{15}{c}{$G$=3}\\
\cline{1-18}\cline{19-36}\\
&&&\multicolumn{5}{c}{General}&\multicolumn{5}{c}{Additive}&\multicolumn{5}{c}{Multiplicative({\bf MIS})}&&&&\multicolumn{5}{c}{General}&\multicolumn{5}{c}{Additive}&\multicolumn{5}{c}{Multiplicative({\bf MIS})}\\\cline{4-8}\cline{9-13}\cline{14-18}\cline{22-26}\cline{27-31}\cline{32-36}\\
 $N$ & $T$ & $\tau$ & $\wh \btheta$&$\wh \bnu$&$\wh \bgamma$&$\wh \varrho_1$&$\wh \varrho_2$ & $\wh \btheta$&$\wh \bnu$&$\wh \bgamma$&$\wh \varrho_1$&$\wh \varrho_2$ & $\wh \btheta$&$\wh \bnu$&$\wh \bgamma$&$\wh \varrho_1$&$\wh \varrho_2$ & $N$ & $T$ & $\tau$ & $\wh \btheta$&$\wh \bnu$&$\wh \bgamma$&$\wh \varrho_1$&$\wh \varrho_2$ & $\wh \btheta$&$\wh \bnu$&$\wh \bgamma$&$\wh \varrho_1$&$\wh \varrho_2$ & $\wh \btheta$&$\wh \bnu$&$\wh \bgamma$&$\wh \varrho_1$&$\wh \varrho_2$\\
 &&&&&&(\%)&(\%)&&&&(\%)&(\%)&&&&(\%)&(\%)& &&&&&&(\%)&(\%)&&&&(\%)&(\%)&&&&(\%)&(\%)\\
\midrule
100 & 50 & 0.1 & 6.1 & 1.9 & 1.0 & 2.8 & 6.8 & 5.1 & 1.9 & 1.0 & 2.7 & 6.7 & 8.9 & 2.0 & 1.0 & 3.2 & 8.3 & 100 & 100 & 0.1 & 14.3 & 2.2 & 1.1 & 9.6 & 19.2 & 8.8 & 2.2 & 1.1 & 9.5 & 17.4 & 17.4 & 2.3 & 1.1 & 9.7 & 20.6\\
 &  & 0.3 & 7.6 & 2.5 & 1.3 &  &  & 6.3 & 2.5 & 1.3 &  &  & 10.7 & 2.6 & 1.4 &  &  &  &  & 0.3 & 17.8 & 3.0 & 1.5 &  &  & 11.3 & 3.0 & 1.5 &  &  & 19.0 & 3.1 & 1.5 &  & \\
 &  & 0.5 & 7.8 & 2.6 & 1.3 &  &  & 6.5 & 2.6 & 1.3 &  &  & 11.2 & 2.8 & 1.4 &  &  &  &  & 0.5 & 19.2 & 3.2 & 1.7 &  &  & 12.5 & 3.2 & 1.7 &  &  & 19.8 & 3.4 & 1.6 &  & \\
 &  & 0.7 & 6.8 & 2.2 & 1.1 &  &  & 5.6 & 2.3 & 1.1 &  &  & 11.0 & 2.5 & 1.2 &  &  &  &  & 0.7 & 16.8 & 2.9 & 1.4 &  &  & 10.7 & 2.9 & 1.4 &  &  & 18.9 & 3.0 & 1.5 &  & \\
 &  & 0.9 & 5.1 & 1.6 & 0.7 &  &  & 4.0 & 1.6 & 0.7 &  &  & 10.9 & 1.8 & 0.9 &  &  &  &  & 0.9 & 13.6 & 2.1 & 1.0 &  &  & 8.6 & 2.0 & 1.0 &  &  & 18.4 & 2.2 & 1.1 &  & \\
\hline
100 & 100 & 0.1 & 3.5 & 1.2 & 0.7 & 0.1 & 0.5 & 2.9 & 1.2 & 0.7 & 0.1 & 0.5 & 7.0 & 1.6 & 0.7 & 0.2 & 0.8 & 100 & 200 & 0.1 & 7.0 & 1.4 & 0.7 & 2.9 & 7.5 & 4.6 & 1.4 & 0.7 & 2.9 & 7 & 14.9 & 1.7 & 0.8 & 3.1 & 12.7\\
 &  & 0.3 & 4.5 & 1.7 & 0.8 &  &  & 3.7 & 1.7 & 0.8 &  &  & 8.8 & 2.1 & 0.9 &  &  &  &  & 0.3 & 8.9 & 2.0 & 0.9 &  &  & 5.9 & 2.0 & 0.9 &  &  & 15.4 & 2.2 & 1.0 &  & \\
 &  & 0.5 & 4.6 & 1.7 & 0.8 &  &  & 3.7 & 1.7 & 0.8 &  &  & 9.5 & 2.1 & 0.9 &  &  &  &  & 0.5 & 9.3 & 2.2 & 1.0 &  &  & 6.4 & 2.1 & 1.0 &  &  & 15.9 & 2.3 & 1.1 &  & \\
 &  & 0.7 & 4.0 & 1.5 & 0.7 &  &  & 3.2 & 1.5 & 0.7 &  &  & 9.8 & 1.9 & 0.8 &  &  &  &  & 0.7 & 8.5 & 2.0 & 0.9 &  &  & 5.7 & 2.0 & 0.9 &  &  & 16.1 & 2.2 & 1.0 &  & \\
 &  & 0.9 & 2.6 & 1.0 & 0.5 &  &  & 2.1 & 1.0 & 0.5 &  &  & 10.0 & 1.5 & 0.7 &  &  &  &  & 0.9 & 6.8 & 1.4 & 0.6 &  &  & 4.7 & 1.4 & 0.6 &  &  & 16.7 & 1.6 & 0.8 &  & \\
\hline
100 & 200 & 0.1 & 2.5 & 0.9 & 0.5 & 0.0 & 0.0 & 2.1 & 0.9 & 0.5 & 0.0 & 0.0 & 6.5 & 1.4 & 0.6 & 0.0 & 0.0 & 200 & 200 & 0.1 & 4.9 & 1.0 & 0.5 & 2.8 & 5.9 & 3.2 & 1.0 & 0.5 & 2.8 & 5.7 & 14.2 & 1.6 & 0.7 & 3.1 & 10.2\\
 &  & 0.3 & 3.2 & 1.2 & 0.6 &  &  & 2.7 & 1.2 & 0.6 &  &  & 8.2 & 1.7 & 0.7 &  &  &  &  & 0.3 & 5.8 & 1.4 & 0.7 &  &  & 4.1 & 1.4 & 0.7 &  &  & 14.4 & 1.8 & 0.8 &  & \\
 &  & 0.5 & 3.2 & 1.3 & 0.6 &  &  & 2.6 & 1.3 & 0.6 &  &  & 9.0 & 1.8 & 0.7 &  &  &  &  & 0.5 & 5.9 & 1.5 & 0.7 &  &  & 4.4 & 1.5 & 0.7 &  &  & 14.7 & 1.9 & 0.8 &  & \\
 &  & 0.7 & 2.8 & 1.1 & 0.5 &  &  & 2.2 & 1.1 & 0.5 &  &  & 9.5 & 1.6 & 0.7 &  &  &  &  & 0.7 & 5.3 & 1.4 & 0.6 &  &  & 3.8 & 1.3 & 0.6 &  &  & 15.0 & 1.7 & 0.8 &  & \\
 &  & 0.9 & 1.9 & 0.7 & 0.3 &  &  & 1.4 & 0.7 & 0.3 &  &  & 9.8 & 1.4 & 0.6 &  &  &  &  & 0.9 & 4.3 & 1.0 & 0.4 &  &  & 3.1 & 1.0 & 0.4 &  &  & 15.8 & 1.4 & 0.7 &  & \\
\hline
200 & 200 & 0.1 & 1.6 & 0.6 & 0.3 & 0.0 & 0.0 & 1.3 & 0.6 & 0.3 & 0.0 & 0.0 & 6.3 & 1.5 & 0.5 & 0.0 & 0.1 & 200 & 400 & 0.1 & 2.8 & 0.7 & 0.4 & 0.3 & 1.4 & 1.9 & 0.7 & 0.4 & 0.3 & 1.3 & 13.8 & 1.5 & 0.6 & 0.4 & 5.8\\
 &  & 0.3 & 2.1 & 0.9 & 0.4 &  &  & 1.7 & 0.8 & 0.4 &  &  & 7.7 & 1.7 & 0.6 &  &  &  &  & 0.3 & 3.5 & 1.0 & 0.5 &  &  & 2.4 & 1.0 & 0.5 &  &  & 13.8 & 1.5 & 0.6 &  & \\
 &  & 0.5 & 2.1 & 0.9 & 0.4 &  &  & 1.8 & 0.9 & 0.4 &  &  & 8.3 & 1.7 & 0.7 &  &  &  &  & 0.5 & 3.6 & 1.0 & 0.5 &  &  & 2.6 & 1.0 & 0.5 &  &  & 14.1 & 1.5 & 0.7 &  & \\
 &  & 0.7 & 1.8 & 0.7 & 0.3 &  &  & 1.5 & 0.7 & 0.3 &  &  & 8.7 & 1.6 & 0.7 &  &  &  &  & 0.7 & 3.2 & 0.9 & 0.4 &  &  & 2.3 & 0.9 & 0.4 &  &  & 14.6 & 1.4 & 0.6 &  & \\
 &  & 0.9 & 1.2 & 0.6 & 0.2 &  &  & 1.0 & 0.5 & 0.2 &  &  & 9.1 & 1.6 & 0.7 &  &  &  &  & 0.9 & 2.4 & 0.7 & 0.3 &  &  & 1.7 & 0.7 & 0.3 &  &  & 15.5 & 1.3 & 0.6 &  & \\
\Xhline{2\arrayrulewidth}
\end{tabular*}
}
\end{sidewaystable}

\vspace{-1.1em}
\subsection{Estimation Performance with Varying $G$ and $H$}
\label{section:vary-group}
\vspace{-0.5em}

In this section, we evaluate the performance of the proposed method under incorrectly specified group numbers, as well as the accuracy of group number estimation in Section \ref{sec:selectG}.
The true number of groups is fixed at $G_0=H_0 = 3$.
To measure the estimation accuracy under this setting, we use the following metrics.
For the estimates of $\bnu^0$ and $\bgamma^0$, we define $\text{RMSE}_{\text{all},\bnu}=(N B)^{-1} \sum_{i=1}^N \sum_{b=1}^B|\wh \nu_{\wh g_i^{(b)}}^{(b)}-\nu_{g_i^0}^0|$ and $\text{RMSE}_{\text{all},\bgamma}=(N B)^{-1} \sum_{i=1}^N \sum_{b=1}^B\|\wh \bgamma_{\wh g_i^{(b)}}^{(b)}-\bgamma_{g_i^0}^0\|$.
For the estimates of $\btheta^0$, we define $\text{RMSE}_{\text{all},\btheta}=N^{-2} B^{-1} \sum_{i=1}^N \sum_{j=1}^N \sum_{b=1}^B |\wh \theta^{(b)}_{\wh g_i^{(b)} \wh h_j^{(b)}}-\theta^0_{g_i^0 h_j^0}|$ to evaluate the estimation accuracy.
We use $\wh \varrho_1$ and $\wh \varrho_2$ defined in \eqref{def: rho1}
to measure the clustering error rates.
In Table \ref{table: group-select}, for concision, we report the mean of these metrics across the five quantiles.
In addition, we also use the QIC criterion proposed in \eqref{eq:QIC} to select the group number, where we set the tuning parameter as $\lambda_{N T}=N^{1 / 10} T^{-1}\log(T)/(10\min\{\ol n, 10\})$.
To evaluate the effectiveness of this selection method, we compute the group number selection accuracy (GSA) as $\operatorname{GSA}=B^{-1} \sum_{b=1}^B I(\wh G^{(b)}=G_0, \wh H^{(b)}=H_0)$.

From Table \ref{table: group-select}, we observe larger RMSEs when the model is under-fitted ($G=2$ or $H=2$),
which suggests significant model estimation bias caused by under-fitted models.
When $G$ and $H$ are over-specified (i.e., $G>3$ or $H>3$), we observe that
the RMSE and clustering errors are also slightly higher than those under the correctly specified model, which is possibly due to the overfitting effects in this case.
Meanwhile, the GSA values also gradually increase with larger $N$ and $T$, which supports the results in Theorem \ref{thm:number}.
The GSA reaches 100\% when $N$ and $T$ are large enough.
In all cases, the resulting $\text{RMSE}_{\text{all},\cdot}$ is very close to that with fixed $G=G_0=3$ and $H=H_0=3$ when $G$ and $H$ are selected by QIC as $\wh G$ and $\wh H$.
Simulations on the power-law networks produce comparable results, with additional details provided in Section B.4.2 of the supplementary material.

\begin{table}[t]
\caption{Simulation results for the SBM network with varying $G$' s.}\label{table: group-select}
\centering
{\fontsize{7}{6}\selectfont
\setlength{\tabcolsep}{7pt}
	\scalebox{0.95}{
\begin{tabular*}{\textwidth}{@{\extracolsep{\fill}}ccccrrrrrrrrrrrr@{\extracolsep{\fill}}}
\toprule
&&&&\multicolumn{6}{c}{ \textsc{Scenario}1}& \multicolumn{6}{c}{ \textsc{Scenario}2}\\\cline{5-10}\cline{11-16}\\
$N$ & $T$ & $G$ & $H$ &  $\wh \btheta$ & $\wh \bnu$ & $\wh \bgamma$ & GSA & $\wh \varrho_1$ & \multicolumn{1}{r}{$\wh \varrho_2$}&$\wh \btheta$ & $\wh \bnu$ & $\wh \bgamma$ & GSA & $\wh \varrho_1$ &{$\wh \varrho_2$}\\
&&&&\multicolumn{3}{r}{(RMSE$_{\text{all}}\times 10^{-2}$)}&(\%)&(\%)&(\%)&\multicolumn{3}{r}{(RMSE$_{\text{all}}\times 10^{-2}$)}&(\%)&(\%)&(\%)\\
\midrule
100 & 50 & \multicolumn{2}{c}{Oracle} & 3.4 & 1.7 & 0.9 & - & - & - & 3.4 & 1.8 & 1.4 & - & - & -\\
 &  & 2 & 2 & 14.4 & 2.2 & 2.0 & - & 42.6 & 35.8 & 13.9 & 2.6 & 3.7 & - & 38.9 & 29.0\\
 &  & 2 & 3 & 18.5 & 2.4 & 2.0 & - & 44.0 & 37.4 & 12.1 & 2.5 & 3.3 & - & 34.8 & 20.0\\
 &  & 3 & 2 & 12.2 & 2.4 & 1.9 & - & 21.0 & 32.3 & 13.8 & 2.6 & 3.2 & - & 22.9 & 29.3\\
 &  & 3 & 3 & 15.9 & 2.5 & 1.9 & - & 22.5 & 34.5 & 11.8 & 2.6 & 3.1 & - & 20.6 & 20.7\\
 &  & 3 & 4 & 18.3 & 2.6 & 2.0 & - & 24.2 & 35.4 & 14.5 & 2.7 & 3.3 & - & 23.3 & 22.7\\
 &  & 4 & 3 & 17.3 & 3.4 & 2.5 & - & 24.1 & 35.8 & 13.6 & 3.3 & 3.9 & - & 23.2 & 22.8\\
 &  & 4 & 4 & 19.6 & 3.5 & 2.5 & - & 25.4 & 36.5 & 16.1 & 3.5 & 4.0 & - & 25.0 & 24.5\\
 &  & $\wh G$ &$\wh H$ & 15.3 & 2.5 & 1.9 & 70.2 & 22.2 & 33.9 & 12.2 & 2.6 & 3.1 & 78.6 & 21.9 & 20.7\\
\hline
100 & 100 & \multicolumn{2}{c}{Oracle} & 2.4 & 1.2 & 0.6 & - & - & - & 2.4 & 1.2 & 1.0 & - & - & -\\
 &  & 2 & 2 & 9.2 & 1.6 & 1.9 & - & 37.6 & 28.1 & 12.4 & 2.2 & 2.5 & - & 27.8 & 33.0\\
 &  & 2 & 3 & 10.6 & 1.6 & 1.9 & - & 38.5 & 25.0 & 11.0 & 2.1 & 2.4 & - & 26.6 & 20.7\\
 &  & 3 & 2 & 8.2 & 1.5 & 1.1 & - & 10.5 & 27.1 & 12.2 & 1.7 & 1.7 & - & 11.1 & 32.3\\
 &  & 3 & 3 & 8.0 & 1.5 & 1.1 & - & 10.1 & 20.5 & 7.8 & 1.6 & 1.5 & - & 7.5 & 14.2\\
 &  & 3 & 4 & 10.0 & 1.5 & 1.1 & - & 10.4 & 22.5 & 10.3 & 1.7 & 1.6 & - & 9.4 & 16.4\\
 &  & 4 & 3 & 9.8 & 2.1 & 1.5 & - & 12.8 & 23.4 & 9.8 & 2.0 & 2.2 & - & 10.8 & 17.7\\
 &  & 4 & 4 & 11.6 & 2.1 & 1.6 & - & 13.5 & 25.1 & 12.1 & 2.2 & 2.3 & - & 12.5 & 19.6\\
 &  & $\wh G$ &$\wh H$ & 8.0 & 1.5 & 1.1 & 94.8 & 10.1 & 20.6 & 7.9 & 1.6 & 1.5 & 92.4 & 7.5 & 14.0\\
\hline
100 & 200 & \multicolumn{2}{c}{Oracle} & 1.7 & 0.9 & 0.4 & - & - & - & 1.7 & 0.9 & 0.7 & - & - & -\\
 &  & 2 & 2 & 7.6 & 1.4 & 1.8 & - & 36.4 & 26.0 & 11.2 & 2.0 & 2.2 & - & 25.6 & 31.0\\
 &  & 2 & 3 & 5.7 & 1.3 & 1.8 & - & 36.2 & 8.9 & 8.3 & 2.0 & 2.1 & - & 24.5 & 13.8\\
 &  & 3 & 2 & 6.6 & 1.0 & 0.6 & - & 3.4 & 25.9 & 10.5 & 1.3 & 1.1 & - & 5.5 & 29.4\\
 &  & 3 & 3 & 3.6 & 0.9 & 0.6 & - & 3.0 & 7.6 & 2.3 & 0.9 & 0.7 & - & 0.7 & 1.6\\
 &  & 3 & 4 & 5.2 & 0.9 & 0.6 & - & 3.1 & 9.9 & 4.1 & 1.0 & 0.8 & - & 1.3 & 2.7\\
 &  & 4 & 3 & 4.5 & 1.3 & 0.9 & - & 4.7 & 8.8 & 3.2 & 1.3 & 1.2 & - & 1.4 & 2.4\\
 &  & 4 & 4 & 6.2 & 1.3 & 0.9 & - & 5.3 & 11.7 & 5.0 & 1.3 & 1.2 & - & 2.2 & 3.6\\
 &  & $\wh G$ &$\wh H$ & 3.6 & 0.9 & 0.6 & 100.0 & 3.0 & 7.6 & 2.3 & 0.9 & 0.7 & 100.0 & 0.7 & 1.6\\
\hline
200 & 200 & \multicolumn{2}{c}{Oracle} & 1.1 & 0.6 & 0.3 & - & - & - & 1.2 & 0.6 & 0.5 & - & - & -\\
 &  & 2 & 2 & 7.6 & 1.2 & 1.7 & - & 28.1 & 26.9 & 12.1 & 2.0 & 2.4 & - & 27.8 & 32.2\\
 &  & 2 & 3 & 4.8 & 1.2 & 1.7 & - & 28.0 & 6.1 & 7.4 & 2.0 & 2.2 & - & 26.5 & 11.2\\
 &  & 3 & 2 & 6.7 & 0.7 & 0.4 & - & 2.9 & 26.7 & 10.0 & 1.1 & 0.9 & - & 5.6 & 29.8\\
 &  & 3 & 3 & 2.6 & 0.7 & 0.4 & - & 2.8 & 5.9 & 1.7 & 0.7 & 0.5 & - & 0.9 & 1.3\\
 &  & 3 & 4 & 4.2 & 0.7 & 0.4 & - & 2.9 & 7.4 & 3.3 & 0.7 & 0.6 & - & 1.3 & 1.9\\
 &  & 4 & 3 & 3.3 & 1.0 & 0.8 & - & 2.9 & 6.4 & 2.3 & 1.0 & 1.1 & - & 1.0 & 1.4\\
 &  & 4 & 4 & 4.8 & 1.0 & 0.8 & - & 3.2 & 8.2 & 3.7 & 1.0 & 1.1 & - & 1.6 & 2.2\\
 &  & $\wh G$ &$\wh H$ & 2.6 & 0.7 & 0.4 & 100.0 & 2.8 & 5.9 & 1.7 & 0.7 & 0.5 & 100.0 & 0.9 & 1.3\\
\hline
200 & 400 & \multicolumn{2}{c}{Oracle} & 0.8 & 0.4 & 0.2 & - & - & - & 0.8 & 0.4 & 0.3 & - & - & -\\
 &  & 2 & 2 & 7.1 & 1.1 & 1.6 & - & 28.0 & 26.4 & 12.0 & 1.9 & 2.2 & - & 27.0 & 31.9\\
 &  & 2 & 3 & 3.9 & 1.1 & 1.6 & - & 28.0 & 1.6 & 6.5 & 1.9 & 2.1 & - & 26.1 & 8.7\\
 &  & 3 & 2 & 6.1 & 0.5 & 0.2 & - & 0.4 & 26.4 & 9.0 & 0.9 & 0.6 & - & 3.2 & 28.1\\
 &  & 3 & 3 & 1.1 & 0.4 & 0.2 & - & 0.3 & 1.4 & 0.9 & 0.4 & 0.3 & - & 0.1 & 0.1\\
 &  & 3 & 4 & 2.1 & 0.4 & 0.2 & - & 0.3 & 1.8 & 1.9 & 0.4 & 0.3 & - & 0.1 & 0.2\\
 &  & 4 & 3 & 1.5 & 0.6 & 0.5 & - & 0.3 & 1.5 & 1.2 & 0.7 & 0.7 & - & 0.1 & 0.1\\
 &  & 4 & 4 & 2.5 & 0.7 & 0.5 & - & 0.4 & 2.0 & 2.2 & 0.7 & 0.7 & - & 0.1 & 0.2\\
 &  & $\wh G$ &$\wh H$ & 1.1 & 0.4 & 0.2 & 100.0 & 0.3 & 1.4 & 0.9 & 0.4 & 0.3 & 100.0 & 0.1 & 0.1\\
\Xhline{2\arrayrulewidth}
\end{tabular*}
}
}
\end{table}
\vspace{-1.1em}

\subsection{Performance under Misspecified Models}
\vspace{-0.5em}

In addition to the general estimation procedures outlined in Section \ref{sec:model_est}, we propose model-specific estimation procedures for the Additive model and Multiplicative model in Section B.3 of the supplementary material to more finely estimate the parameters.
In this context, we proceed to investigate the robustness of the TGNQ model by studying its performance when the model is misspecified but $G = G_0$ and $H = H_0$.
In \textsc{Scenario} 1, we employ the general, additive, and multiplicative models to estimate the parameters.
Since the data is generated by the additive model, the multiplicative model represents a misspecified model in this case.
The results are summarized in Table \ref{table: mis-add}.
In \textsc{Scenario} 2, we again utilize the general, additive, and multiplicative models to estimate the parameters,
where the additive model is now a misspecified model.
Similar results for \textsc{Scenario} 2 can be found in Section B.4.2 of the supplementary material.

In Scenario 1, as shown in Table \ref{table: mis-add}, we observe that when utilizing the misspecified model to fit the data, the group membership estimation error rate rises slightly but remains acceptable.
However, compared to the specified models, the root mean squared error (RMSE) of the fitted parameter $\btheta^0$ is higher and fails to decrease considerably with rising $N$ and $T$.
From the preceding analyses, the estimation methodology of the TGNQ model demonstrates appreciable robustness across varying scenarios.
Results consistent with those obtained from simulations in \textsc{Scenario} 2 and those on power-law networks are documented in Section B.4.2 of the supplementary material for further reference.

\vspace{-1.1em}
\subsection{Comparison with GNAR Model}\label{section:compare_qgnar}
\vspace{-0.5em}
To further validate the effectiveness of our proposed TGNQ model, we compare it with the quantile version of the Group Network Autoregressive (GNAR \citep{zhu2025simultaneous}) model and abbreviate it as QGNAR, where the model form and the estimation methods are described in detail in Section B.7 of the supplementary material.
We compare the model estimation performance of both models at quantile levels $\tau \in \{0.1, 0.3, 0.5, 0.7, 0.9\}$.
Specifically, the true number of groups is set to $G_0 = H_0 = 2$.
For TGNQ, we set $G = H = 2$, and for QGNAR, we set $G = 2$, and we repeat the experiment $B=500$ times.
We calculate the estimation accuracy of the autoregressive matrix $\B(\tau)$ in \eqref{model1} to evaluate the model performance.
Let $\cS = \{(i,j): a_{ij}\ne 0\}$ collect the network edges, and define
the root mean squared error (RMSE) as
\compressmath
$$
\text{RMSE}_{\cS} = \Big\{{\frac{1}{B|\cS|} \sum_{b=1}^B\sum_{(i,j) \in \cS} \left( \wh{B}_{ij}(\tau) - B_{ij}^0(\tau) \right)^2}\Big\}^{1/2},
$$
where $\B^0(\tau) = (B_{ij}^0(\tau))$ is the true autoregressive coefficient matrix,
and $\wh \B(\tau) = (\wh{B}_{ij}(\tau))$ denotes its estimate.

\begin{table}[t]
\caption{Simulation results for TGNQ and QGNAR models.}\label{table: gnar}
\centering
{\fontsize{8}{9}\selectfont
\setlength{\tabcolsep}{7pt}
	\scalebox{0.95}{
\begin{tabular*}{\textwidth}{@{\extracolsep{\fill}}cccrrrrrrrr@{\extracolsep{\fill}}}
\toprule
&&& \multicolumn{4}{c}{SBM}& \multicolumn{4}{c}{Powerlaw}
\\
&&& \multicolumn{2}{c}{ \textsc{Scenario}1}& \multicolumn{2}{c}{ \textsc{Scenario}2}& \multicolumn{2}{c}{ \textsc{Scenario}1}& \multicolumn{2}{c}{ \textsc{Scenario}2}\\\cline{1-3}\cline{4-5}\cline{6-7}\cline{8-9}\cline{10-11}
$N$ & $T$ & $\tau$ & TGNQ &  QGNAR & TGNQ & QGNAR & TGNQ & QGNAR & TGNQ & QGNAR\\
&&&\multicolumn{2}{r}{(RMSE$_{\cS}\times 10^{-2}$)}&\multicolumn{2}{r}{(RMSE$_{\cS}\times 10^{-2}$)}&\multicolumn{2}{r}{(RMSE$_{\cS}\times 10^{-2}$)}&\multicolumn{2}{r}{(RMSE$_{\cS}\times 10^{-2}$)}\\
\midrule
100 & 50 & 0.1 & 1.38 & 3.03 & 1.51 & 2.28 & 1.62 & 3.45 & 1.65 & 2.57 \\
   &  & 0.3 & 1.45 & 3.05 & 1.69 & 2.41 & 1.71 & 3.48 & 1.83 & 2.70 \\
   &  & 0.5 & 1.48 & 3.07 & 1.76 & 2.50 & 1.72 & 3.50 & 1.92 & 2.81 \\
   &  & 0.7 & 1.45 & 3.08 & 1.77 & 2.59 & 1.69 & 3.52 & 1.94 & 2.90 \\
   &  & 0.9 & 1.40 & 3.09 & 1.76 & 2.73 & 1.65 & 3.55 & 1.94 & 3.03 \\ \hline
  100 & 100 & 0.1 & 0.37 & 3.02 & 0.57 & 2.26 & 0.62 & 3.44 & 0.76 & 2.55 \\
   &  & 0.3 & 0.42 & 3.04 & 0.63 & 2.38 & 0.67 & 3.45 & 0.83 & 2.68 \\
   &  & 0.5 & 0.42 & 3.06 & 0.66 & 2.48 & 0.68 & 3.47 & 0.86 & 2.79 \\
   &  & 0.7 & 0.39 & 3.07 & 0.65 & 2.57 & 0.65 & 3.49 & 0.87 & 2.89 \\
   &  & 0.9 & 0.33 & 3.08 & 0.65 & 2.72 & 0.59 & 3.53 & 0.87 & 3.02 \\ \hline
  100 & 200 & 0.1 & 0.17 & 3.02 & 0.18 & 2.25 & 0.24 & 3.43 & 0.27 & 2.55 \\
   &  & 0.3 & 0.21 & 3.02 & 0.22 & 2.37 & 0.29 & 3.43 & 0.30 & 2.67 \\
   &  & 0.5 & 0.21 & 3.05 & 0.22 & 2.46 & 0.28 & 3.45 & 0.30 & 2.77 \\
   &  & 0.7 & 0.18 & 3.06 & 0.20 & 2.56 & 0.25 & 3.47 & 0.29 & 2.88 \\
   &  & 0.9 & 0.12 & 3.07 & 0.15 & 2.71 & 0.19 & 3.52 & 0.25 & 3.01 \\ \hline
  200 & 200 & 0.1 & 0.13 & 3.32 & 0.16 & 2.55 & 0.19 & 3.69 & 0.24 & 2.81 \\
   &  & 0.3 & 0.16 & 3.32 & 0.18 & 2.67 & 0.22 & 3.68 & 0.27 & 2.93 \\
   &  & 0.5 & 0.16 & 3.35 & 0.19 & 2.78 & 0.21 & 3.70 & 0.27 & 3.05 \\
   &  & 0.7 & 0.14 & 3.36 & 0.17 & 2.89 & 0.20 & 3.71 & 0.27 & 3.17 \\
   &  & 0.9 & 0.09 & 3.38 & 0.14 & 3.04 & 0.16 & 3.73 & 0.25 & 3.33 \\
\Xhline{2\arrayrulewidth}
\end{tabular*}
}
}
\end{table}

Table \ref{table: gnar} summarizes the average $\text{RMSE}_\cS$ values of the two methods across different $(N, T, \tau)$ settings.
As the sample size increases, the TGNQ model's error steadily decreases, indicating good convergence. In contrast, the GNAR model shows little improvement, suggesting substantial estimation bias. This highlights that when the true model has a two-way structure, using a one-way approach such as QGNAR can lead to significant bias.

\vspace{-1.1em}
\section{Real Data Example}\label{section: real}

\vspace{-0.5em}
 
\subsection{Data Description}
\vspace{-0.5em}

This study conducts an empirical analysis based on user behavioral data from Sina Weibo, a widely used Chinese social media platform.
Specifically, the sample consists of 355 active users observed over 78 consecutive days (January 1 to March 19, 2014),
with network relationships defined by mutual following behaviors.

\begin{figure}[htpb!]
\centering
\includegraphics[width=0.9\textwidth]{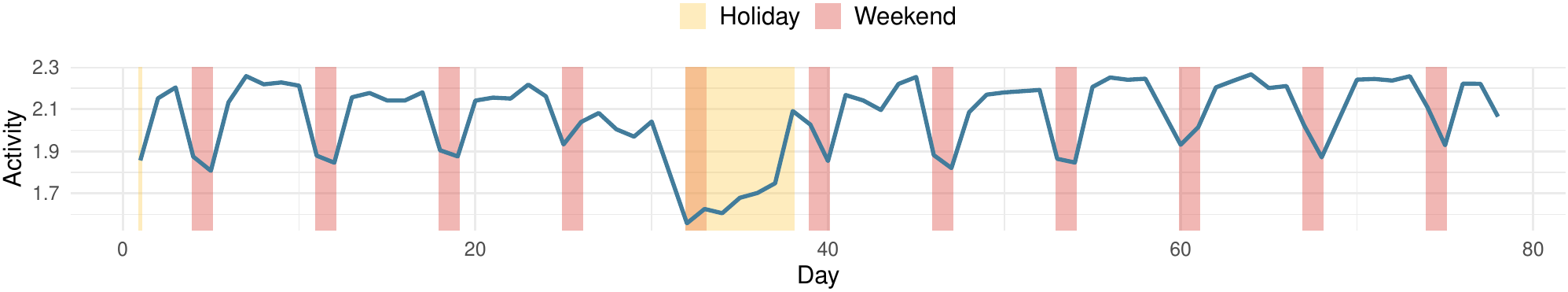}
 \caption{Time series of cross-sectional averages of activity across users.}
\label{fig:y_plot}
\end{figure}

In our analysis, we focus on user activity as the outcome variable.
Following \cite{zhu2017network}, user activity is measured as the log-transformed posting frequency, defined as $Y_{it}=\log(1+Z_{it})$, where $Z_{it}$ represents the number of posts by user $i$ on day $t$.
Figure \ref{fig:y_plot} shows the time series of daily average responses over $1\le i\le N$, revealing cyclical patterns with consistently lower activity on weekends and national holidays compared to weekdays.
Accordingly, we introduce two time-varying control variables: Weekend (1 for Saturday-Sunday, 0 for weekdays) and Holiday (1 for national holidays, 0 otherwise) to capture the temporal effects on user behaviors.

\vspace{-1em}
\vspace{-1em}
To explore heterogeneity in user activity, we collect seven user characteristics, including
Gender (1 for male, 0 for female),
Tenure (years since account creation),
Beijing (1 for Beijing-based users, 0 otherwise),
Shanghai (1 for Shanghai-based users, 0 otherwise),
Description (logarithm of character count of profile descriptions),
Weibo (logarithm of total post volume),
and Public (1 for verified public accounts, 0 otherwise).

\vspace{-1.1em}
\subsection{Model Estimation Results}\label{section: real-result}
\vspace{-0.5em}

\begin{figure}[htpb!]
\centering
 \includegraphics[width=0.5\textwidth]{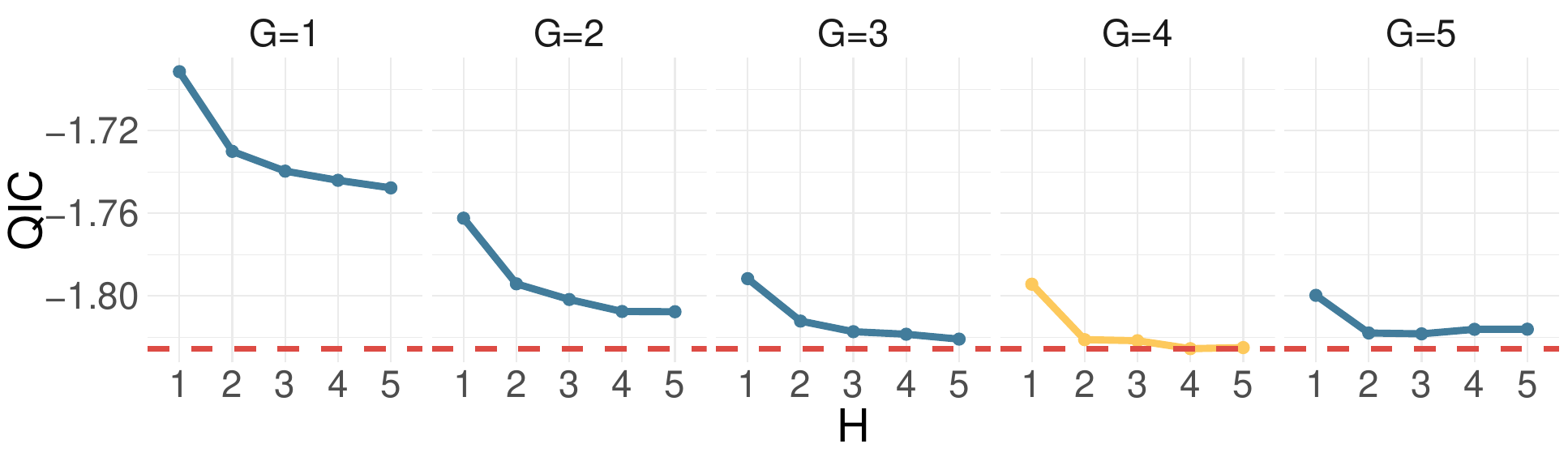}
 \caption{\small {QIC values for $1\leq G \leq 5$ and $1\leq H \leq 5$.}}
\label{fig:selectGH}
\end{figure}

To apply the TGNQ model, we first select the number of groups $G$ and $H$.
The model is trained on the first 71 days of data with $\lambda_{NT} = N^{1/10} T^{-1}\log(T)/(10\min\{\ol{n}, 10\})$ as in the simulation study.
As shown in Figure \ref{fig:selectGH}, the QIC is minimized at $\wh G=4$ and $\wh H=4$.
For $\wh G=4$, the QIC value decreases sharply when $H \leq 2$, while the change becomes more gradual when $H > 2$.
This occurs because, when group numbers are over-specified, the estimation consistency is still guaranteed by Theorem \ref{thm:consistency}, resulting in smaller decreases in empirical loss.
We further evaluate models with $\wh G=4$ and $\wh H=2,3,4,5$ by predicting outcomes using the final week of data as the validation set,
finding the lowest prediction loss at $\wh H=3$ (see Section B.8 of the supplementary material).
Consequently, we select $\wh G=4$ and $\wh H=3$ and refit the model using the full dataset.
Before presenting the estimation results, we validate our method via simulation experiments based on the estimated network structure and parameters; see Section B.8 of the supplementary material.

 \begin{figure}[htpb!]
 \centering
 \includegraphics[width=0.31\textwidth]{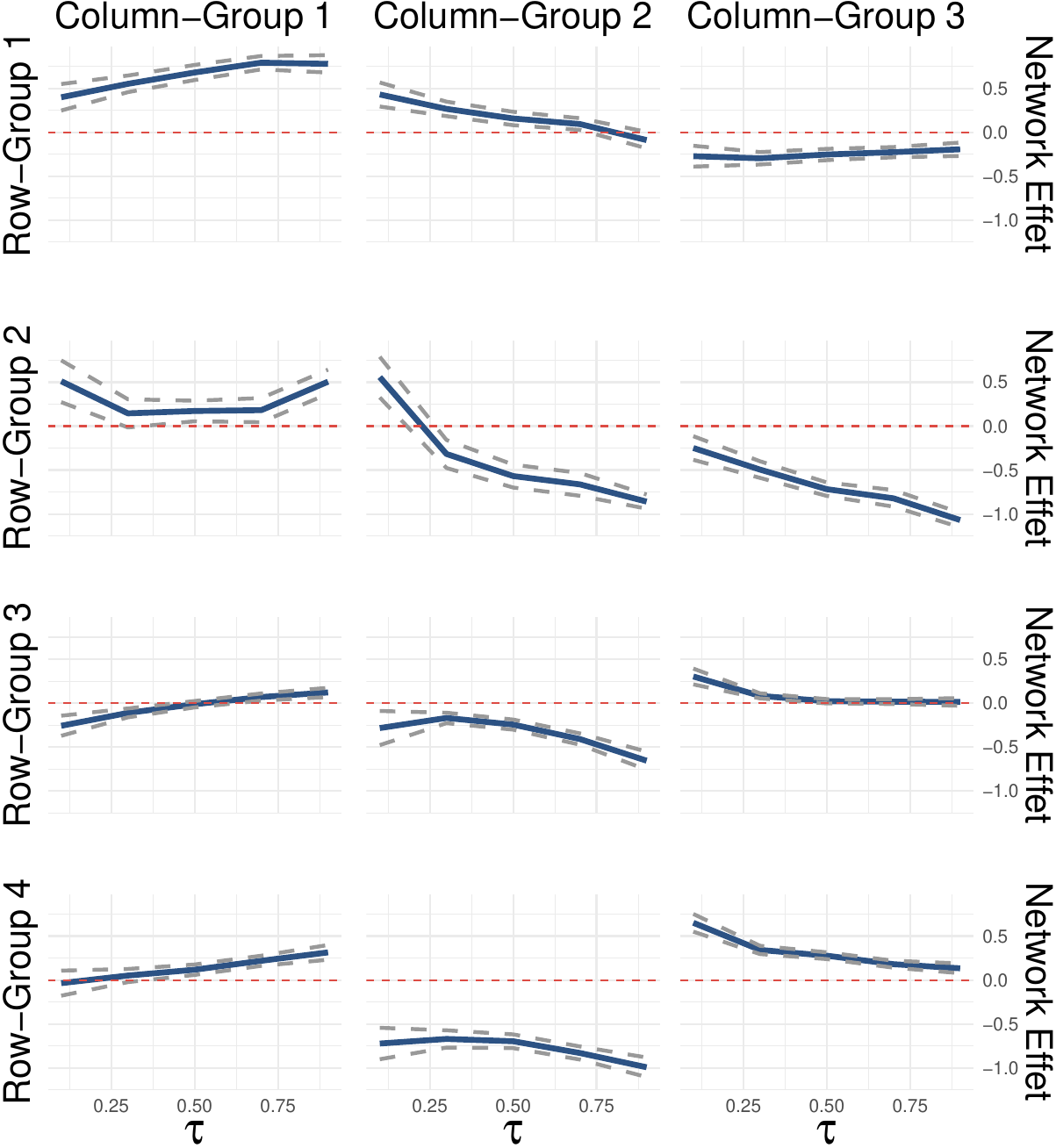}
 \includegraphics[width=0.3\textwidth]{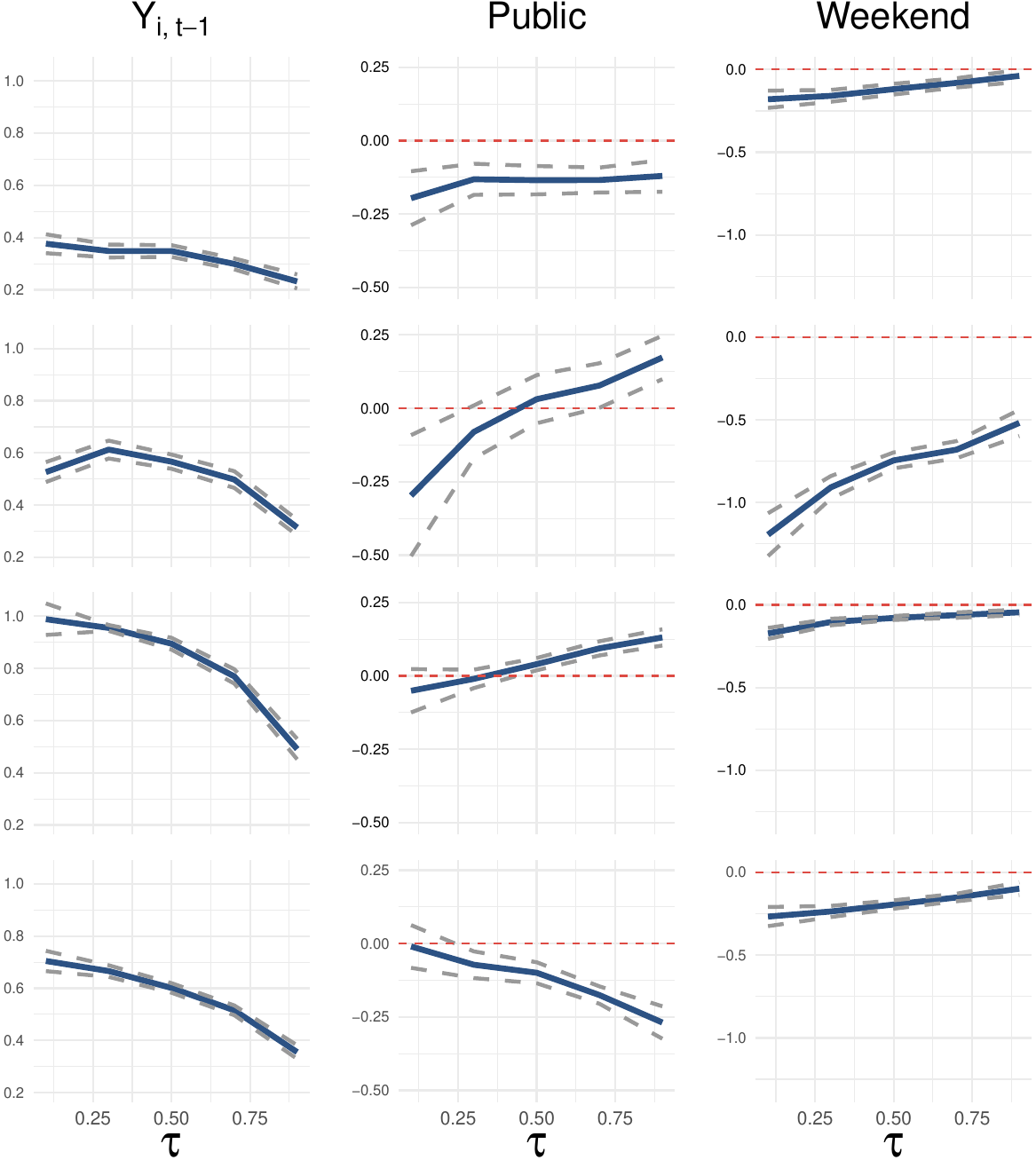}
 \includegraphics[width=0.2\textwidth]{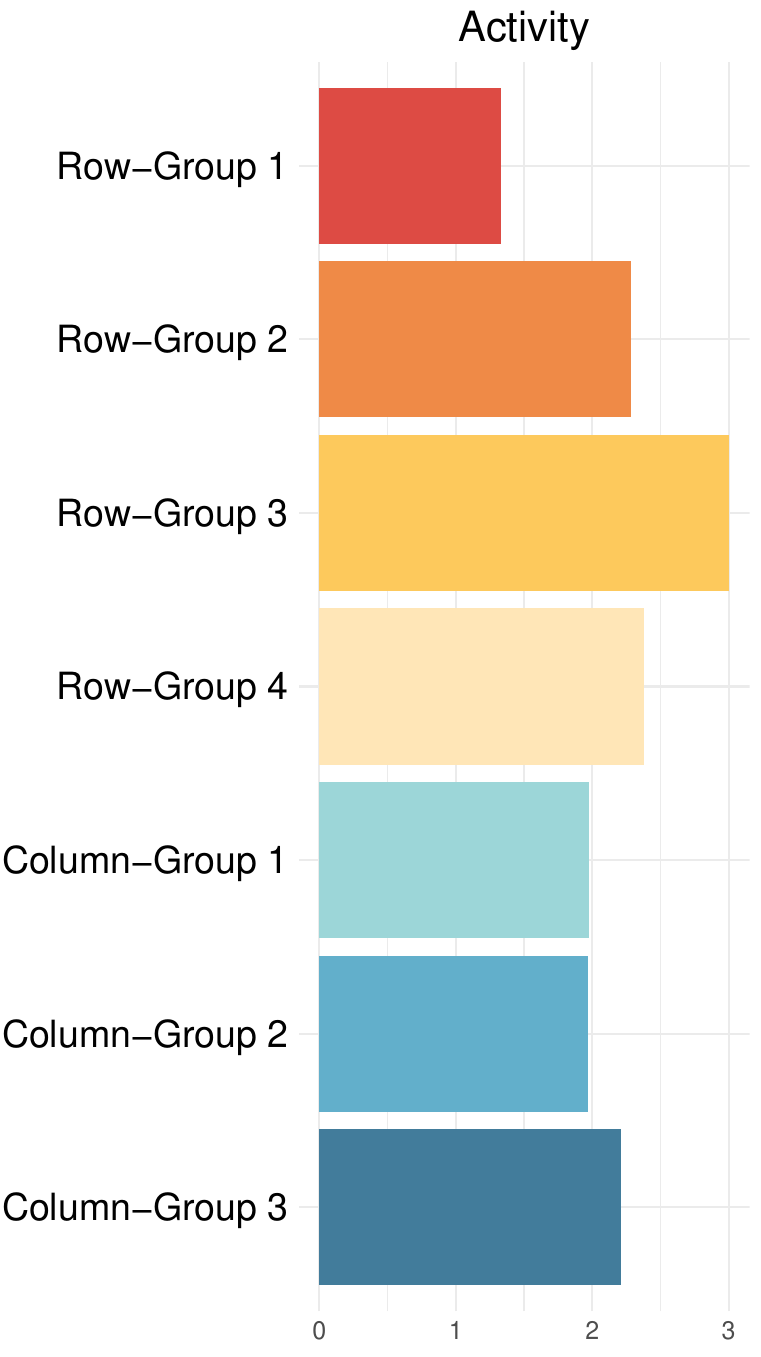}
 \caption{\small {Parameter estimates across $\tau$. Left panel: network effects with 95\% confidence interval bounds; middle panel: autoregressive and covariate effects with 95\% confidence interval bounds;
 right panel: Groupwise Means of Response.}}
\label{fig:para_real}
\end{figure}

Figure \ref{fig:para_real} presents the group-level parameter estimates, highlighting clear group-level heterogeneity.
From a transmission perspective, as quantile $\tau$ increases, Column-Group 1 shows a rising and mainly positive network effect.
This indicates that Column-Group 1 possesses a stronger capability for activity diffusion.
In contrast, Column-Groups 2 and 3 exhibit decreasing network effects with increasing $\tau$,
displaying complex patterns of influence.
For example, Column-Group 2 exhibits a positive network effect on Row-Group 1, but relatively negative effects on Row-Groups 2--4.
Building on these findings, Table \ref{tb:accounts} presents public accounts with over 350,000 followers.
Specifically, Column-Group 1 includes broad-reach accounts such as ``Phoenix TV'' and ``China News Service'',
whose influence is more readily received and amplified at higher quantiles.
In contrast, Column-Groups 2--3 contain more specialized, niche-oriented accounts that may yield diminished marginal gains in cross-group influence at higher quantiles.
Notably, Column-Group 2 consistently has significant negative effects on the Row-Groups 3 and 4, indicating a ``crowding-out effect''.
By contrast, the influence of Column-Group 3 is generally moderate or positive.
In addition,
Row-Group 3 is notably self-driven,
with the highest response mean and strongest autoregressive effects, but minimal network influence.
This indicates their activity is sustained mainly by intrinsic motivation.

\vspace{-1em}
\begin{table}[h!]
\caption{Verified public accounts with over 350,000 followers.}\label{tb:accounts}
{\fontsize{10}{12}\selectfont
\setlength{\tabcolsep}{4pt}
\begin{tabular}{ccl}
\toprule
Column-Group       & Account Name               & Description                                                       \\ \hline
\multirow{2}{*}{1} & Phoenix TV                 & Global Chinese-language satellite TV channel                      \\ \cline{2-3}
                   & China News Service         & National-level media                                              \\ \hline
\multirow{3}{*}{2} & YICAI                      & Financial media group                                             \\ \cline{2-3}
                   & Enterprise Weibo Assistant & Official account for enterprise Weibo operations                  \\ \cline{2-3}
                   & Japan Trend Frontline      & Japanese information Weibo                                        \\ \hline
\multirow{4}{*}{3} & Otaku Bar                  & Self-media creator focusing on humorous content \\ \cline{2-3}
                   & Hot Movies                 & Movie reviews                                                     \\ \cline{2-3}
                   & Sohu Video                 & Comprehensive video website                                       \\ \cline{2-3}
                   & Durex                      & Globally renowned sexual health brand                             \\ \Xhline{2\arrayrulewidth}
\end{tabular}
}
\end{table}
\vspace{-1.5em}
Furthermore, autoregressive effects are significantly positive across all groups, but decrease in strength at higher activity quantiles.
Covariate effects also reveal heterogeneity: the Public effect in Row-Group 3 is higher than in Row-Group 1,
while Row-Groups 2 and 4 display opposite trends across quantiles, reflecting varying levels of public influence.
Lastly, Weekend effects are universally negative and most pronounced in Row-Group 2, consistent with lower weekend activity observed in the descriptive analysis.

\vspace{-0.7em}
 \begin{figure}[htpb!]
 \centering
    \includegraphics[width=0.7\textwidth]{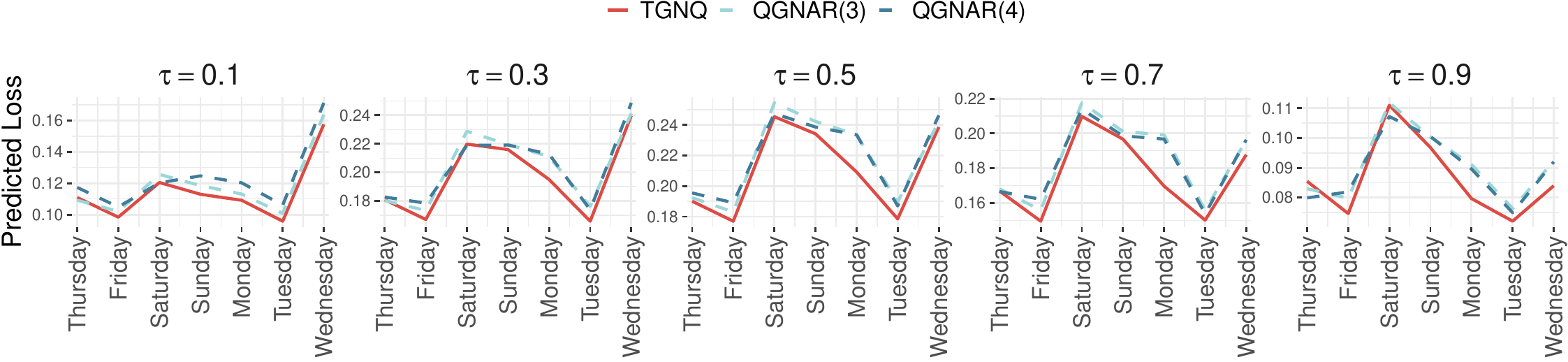}
 \caption{Comparison of Predicted Loss Across Quantiles for Different Models}
\label{fig:prediction}
\end{figure}
\vspace{-1.5em}

To further validate the TGNQ model's effectiveness, we conduct comparative analyses against QGNAR (with only one-way group structure), considering configurations with $G=3$ (9 network parameters) and $G=4$ (16 network parameters), while TGNQ requires 12 network parameters.
Specifically, models are trained on the first 71 days of data.
Results show that TGNQ consistently achieves a better fit across all quantiles (see Section B.8 of the supplementary material for details).
Subsequently, we predict user activity for the last 7 days and compute prediction losses at five quantile levels.
Figure \ref{fig:prediction} shows that TGNQ consistently achieves lower prediction losses across most quantiles and time points compared to both QGNAR configurations.
These results confirm that the two-way grouping structure enhances both fitting accuracy and predictive performance.

\vspace{-1.1em}

\section{Conclusion}\label{section:conclusion}

\vspace{-0.5em}
This paper introduces a two-way grouped network quantile panel data model, capturing cross-sectional and temporal dependence in networks.
The novel structure allows distinct ``receiving'' and ``influencing'' group memberships for each unit.
We develop an efficient estimation algorithm and establish key asymptotic properties, including consistency and asymptotic normality under certain conditions.
Additionally, we propose a data-driven criterion for group number selection.
Simulations demonstrate strong finite-sample performance,
while an application to Sina Weibo data showcases the model's practical utility.
Building on these foundations, we envision several promising avenues for future research.
These include exploring time-varying network structures, investigating semi/non-parametric specifications,  conducting theoretical analyses with growing covariate numbers, and developing robust inference methods under network structure uncertainty.
In addition, it would be interesting to 
investigate statistical inference methods based on the pre-refinement estimator.
Furthermore, improving the computational efficiency and establishing theoretical properties of the discrete optimization procedure for larger networks remain an important challenge.
Such extensions will further enhance the model's applicability and deepen our understanding of empirical research contexts.

\vspace{-2em}
\section*{Acknowledgments}
\vspace{-1.1em}

Wenyang Liu and Xuening Zhu's research is supported by the National Natural Science Foundation of China (nos. 72573038, 12331009), MOE Laboratory for National Development and Intelligent Governance, Fudan University.

\vspace{-2em}
\section*{Conflict of Interest}
\vspace{-1.1em}

The authors report there are no competing interests to declare.

\bibliographystyle{asa}
\begingroup
\setstretch{0.9}
\setlength{\bibsep}{1.1pt}
\fontsize{9.5}{9}\selectfont
\bibliography{xuening}
\endgroup
\end{document}